\documentstyle[amsmath,psfig]{mn}

\def \msun{\text{M}_{\odot}}

\title[Quasars: a toroidal black hole interpretation]
{Quasars: a supermassive rotating toroidal black hole interpretation}

\author[R. J. Spivey]
{Robin J. Spivey\thanks{email: torus@physicist.net} }

\date{Accepted 2000 March 24. Received 2000 March 15;
in original form 1999 March 18}

\pagerange{\pageref{firstpage}--\pageref{lastpage}}

\pubyear{2000}
\volume{316}
\pagerange{856-874}
\journal{Mon. Not. R. Astron. Soc. {\bf 316}, 856-874 (2000)}

\begin{document}

\maketitle

\label{firstpage}

\begin{abstract}
  A supermassive rotating toroidal black hole (TBH) is proposed as the
  fundamental component of quasars and other jet-producing active
  galactic nuclei. Rotating protogalaxies gather matter from the central
  gaseous region leading to the birth of massive toroidal stars, the
  internal nuclear reactions of which proceed very rapidly. Once the
  nuclear fuel is spent, gravitational collapse produces a slender TBH
  remnant. Transitory electron and neutron degeneracy stabilised collapse
  phases, although possible, are unlikely owing to the large masses
  involved thus these events are typically the first supernovae of the
  host galaxies. Given time, the TBH mass increases through continued
  accretion by several orders of magnitude, the event horizon swells
  whilst the central aperture shrinks. The difference in angular
  velocities between the accreting matter and the TBH induces a magnetic
  field that is strongest in the region of the central aperture and
  innermost ergoregion. Due to the presence of negative energy states
  when such a gravitational vortex is immersed in an electromagnetic
  field, circumstances are near ideal for energy extraction via
  non-thermal radiation including the Penrose process and superradiant
  scattering. This establishes a self-sustaining mechanism whereby the
  transport of angular momentum away from the quasar by relativistic
  bi-directional jets reinforces both the modulating magnetic field and
  the TBH/accretion disc angular velocity differential. Continued
  mass-capture by the TBH results in contraction of the central aperture
  until the TBH topology transitions to being spheroidal, extinguishing
  quasar behaviour. Similar mechanisms may be operating in microquasars,
  supernovae and sources of recurrent gamma-ray bursts when neutron
  density or black hole tori arise. In certain circumstances, long-term
  TBH stability can be maintained by a negative cosmological constant,
  otherwise the classical topology theorems must somehow be circumvented.
  Preliminary evidence is presented that Planck-scale quantum effects may
  be responsible.
\end{abstract}


\section{INTRODUCTION}

    The commonly cited method of producing ultra-relativistic 
bi-directional jets as observed in quasars is the mechanism described by 
Blandford \& Znajek (1978), whereby magnetic field lines thread the poles 
of a rotating BH as they descend towards the event horizon. Rotational 
energy may be extracted from the BH by this technique which is ejected in 
the form of radiation and matter travelling at high velocity along the 
BH's spin axis. Critical assessments by Ghosh \& Abramowicz (1997) and 
Livio, Ogilvie \& Pringle (1999) suggest that the role of the 
Blandford-Znajek mechanism has been generally overestimated and 
inadequately accounts for the larger double radio lobe structures. 
Numerical simulations indicate that the observed gamma ray energy release 
along quasar jets is four orders of magnitude more energetic than the 
Blandford-Znajek mechanism predicts. Issues that are difficult to 
reconcile with this model are the variability of jet dispersion angles, 
the finite quasar lifetime and the multiplicity of red-shifts in the very 
metallic absorption spectra. Any viable alternative model must 
simultaneously cater for all features.

    Speculation concerning the fundamental processes governing quasars 
invariably involves discussion of compact massive central bodies, the 
consensus being that these are rotating spheroidal BHs of mass 
$10^6$--$10^9\msun$. Profiles of stellar orbital velocities within AGN 
haloes lend weight to the premise that massive objects reside at the 
galactic nuclei. Inactive galactic nuclei (IGN) have yielded comparable 
velocity profiles, suggesting that the masses of AGN and IGN are similar, 
if not identical. The vastly differing activity levels could signify an 
inherent defect with current AGN models. Little theoretical progress seems 
to have been made over past decades towards a full explanation for these 
exceedingly energetic phenomena. Closer inspection and revision of 
existing theories may be necessary to attain a consistent understanding of 
quasars and AGN. The crucial test of any theory is the correspondence 
between predictions and observations, it is argued that existing models 
are struggling in this respect. The purpose of this discussion is to 
advocate a new model and describe how, with relatively minor theoretical 
embellishments, compelling explanations for AGN and anisotropic gamma-ray 
bursts accompanying core-collapse supernovae can be developed. Possible 
mechanisms responsible for originating, accelerating and collimating jets 
are then discussed.

    The well known black hole uniqueness theorems rest on the classical 
theory of general relativity. It is thought that any non-distorted and 
asymptotically flat black hole spacetime can be represented by the 
Kerr-Newman set of solutions. But, while an ultimate theory remains 
elusive, one cannot be entirely sure of the validity of the uniqueness 
theorems. Hence, it is important to explore possibilities beyond those 
anticipated by purely classical calculations. Small departures from 
classical physics, e.g. of quantum mechanical origin, might lead to 
profound macroscopic changes, even to the extent that the topology of 
black hole horizons can be altered. Similarities between a rotating 
toroidal black hole (TBH) at the galactic centre accreting matter from its 
surroundings will be compared with observational evidence from quasars, 
Seyferts, BL Lacertae and blazars, which have long been suspected to be 
manifestations of the same underlying astrophysical phenomena. Attention 
will be paid to the formation of such a TBH, its long-term stability in 
our universe, jet production and its evolution with time. The possibility 
is examined that quasars may have been present at some stage of almost 
every galaxy's development in the early universe. A diagram encapsulating 
the life-cycles of toroidal BHs, in qualitative agreement with those of 
quasars is finally presented.

\section{ROTATING TOROIDAL BLACK HOLES AND THEIR FORMATION}

    It is proposed that the central component of the quasar mechanism is a 
rapidly rotating black hole with a toroidal event horizon. First, the 
possible embryonics of formation are addressed. The constituent stars of 
most observable galaxies are concentrated in the plane of galactic 
rotation. Direct observations of the cosmic microwave background by the 
COBE satellite indicates that matter was very evenly distributed 
throughout the cosmos in earlier times. Thereafter, on the scale of 
inter-galactic distances, matter must have collapsed under the action of 
gravity, triggering the emergence of protogalaxies composed of low-density 
hydrogen and helium gas. Because most galaxies are observed to rotate, 
these protogalaxies would generally have possessed angular momentum.

    Protogalactic gas clouds then draw towards the plane of rotation. 
Molecules have random velocities but collisions are relatively infrequent 
owing to low particle densities. Those with small velocities tend to 
accumulate at the galactic centre due to net gravitational attraction. 
However, these particles gain kinetic energy as they proceed towards the 
centre, preventing the majority from occupying orbits confined to the 
centremost regions of the galaxy. Instead they tend to cluster in 
elliptical orbits of larger radii resulting in a relative underabundance 
of particles at the core, see Fig.~1, curve labelled $t=1$. For a given 
instant in time, the density within the galactic plane is at a local 
minimum at the centre, increases with radius to a maximum and thereafter 
tapers off. As time progresses, the gas distribution becomes more 
pronounced and collisions between molecules more frequent. The evolution 
of the distribution in this scenario is qualitatively depicted by the 
series of gas density curves in Fig.~1. The curve labelled $t=4$ 
represents what may be identified as a toroidal gas cloud. This ultra-low 
metallicity toroidal proto-star continues to condense as more gas 
molecules amass until the density and pressure are sufficient for nuclear 
fusion. The toroidal stars of this model, Fig.~2(a), will exhaust their 
nuclear fuel exceedingly rapidly as regions suitable for fusion occupy a 
greater volumetric fraction of toroidal stars than of spherodial stars.

    By comparison with the upper size and mass limits for spheroidal 
neutron stars, an upper limit can be estimated for the minor radius $R_2$ 
of a neutron torus above which collapse to a TBH will result\footnote{At 
the time of publication, this TBH formation route seemed to me the most 
likely, avoiding the need for an SBH$\to$TBH transition. Since then, I am 
of the opinion that the transition can occur in either direction and that 
the toroidal progenitor is merely an interesting possibility, not an 
essential ingredient of the model.}. Simplifying assumptions are employed. 
First, the density of neutron degenerate material is assumed to remain 
constant and independent of pressure as for an incompressible fluid. 
Hydrostatic equilibrium is reached whereby the pressure of the fluid 
counteracts gravitational compression at all locations. Newtonian 
approximations will be used to derive the surface gravity. The torus is 
assumed to have a major radius much larger than the minor radius, $R_1\gg 
R_2$, so that an infinitely long cylinder approximation is valid. Rotation 
is  neglected. The gravitational field within a sphere of constant density 
tails off linearly from the surface to the centre, even according to 
general relativity. It is useful to confirm that gravity within an 
infinitely long solid cylinder of constant mass density is linearly 
related to the radial distance from the axis according to a Newtonian 
analysis.

\begin{figure*}
\centerline{\psfig{figure=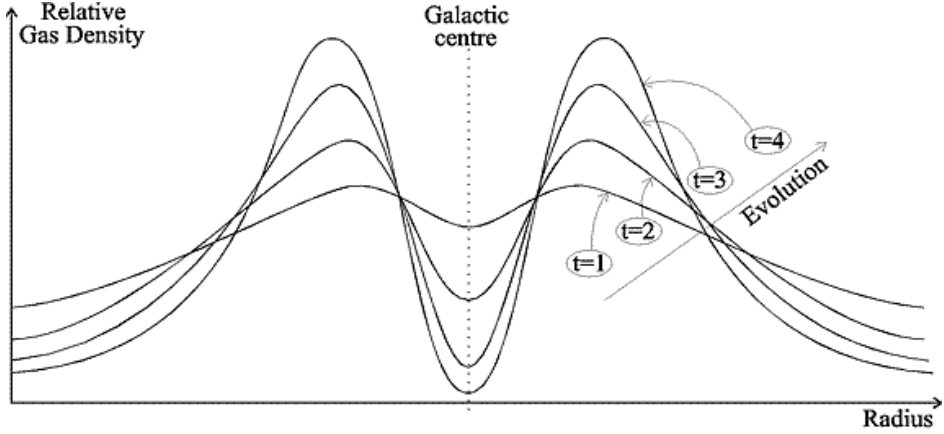,width=0.7\textwidth,angle=0}}
\caption{Evolution of a rotating, initially collisionless dust cloud. 
Axial densities decrease with time since these regions can only 
accommodate particles with both minimum kinetic and gravitational 
potential energies. Otherwise, the particles are drawn to the equatorial 
plane and a toroidal structure inevitably develops. Dissipative processes 
diminish the ellipticity of orbital trajectories.}
\end{figure*}

    For points external to spherically symmetric objects, the gravity is 
known to be equivalent to that of a point particle of equal total mass 
located at the centre of symmetry, regardless of any radial density 
variations. Similarly, the external gravity of an infinitely long 
cylindrically symmetric mass is equivalent to that of a line mass of 
infinite length located on the axis. To prove this, assign an outer radius 
to the cylinder of $R_{\rm\,T}$, a longitudinal coordinate $x$ along the 
cylinder's length and angular coordinate $\vartheta$. The gravitational 
field strength at some radius $a<R_{\rm\,T}$ inside the cylinder with 
longitudinal coordinate $x=0$ and angular coordinate $\vartheta=0$ is 
sought. This location is external to a cylinder of radius $a$ and internal 
to a cylindrical shell of radial thickness $b-a$. Because Newtonian 
gravity obeys the principle of superposition it is first demonstrated that 
the gravitational field vanishes at all points located within an 
infinitesimally thin and infinitely long cylindrical shell with constant 
mass per unit area $\sigma$. Then, by integrating the gravitational 
contribution of internal cylindrical shells, gravity inside an infinite 
homogeneous cylinder is observed to vary linearly with radius, as is the 
familiar variation within homogeneous spheres.

    Let the thin cylindrical shell have radius $b>a$. Integrating the 
radially directed gravitational field contributions of elemental masses of 
constant radius $b$ over the integration variables $x$ and $\vartheta$, an 
expression is obtained of the form:
\begin{align}
g(a)\!=2G\sigma \!\!\int_0^\pi\!\!\!\int_{-\infty}^{+\infty}\!\!\!\!\frac{
b(a-b\cos\vartheta)}{(a^2+b^2-2ab\cos\vartheta+x^2)^{3/2}}\,dx\,d\vartheta
\end{align}

Integrating with respect to $x$ gives:
\begin{multline}
g(a)=2G\sigma\left[\frac{x}{\sqrt{x^2+a^2-2ab\cos\vartheta+b^2}}\right]_{
-\infty}^{+\infty}\\
\times\int_0^\pi\left(\frac{ab-b^2\cos\vartheta}{a^2-2ab\cos\vartheta+b^2}
\right)\,d\vartheta\\
\text{thus,\quad}g(a)=4G\sigma\int_0^{\pi}\frac{ab-b^2\cos\vartheta}{
a^2-2ab\cos\vartheta+b^2}\,d\vartheta
\end{multline}

Splitting the integral in two, integrating with respect to $\vartheta$ and 
recalling that $b>a$ yields:
\begin{align}
g(a)=4\pi G\sigma ab\left(\frac{1}{|b^2-a^2|}-\frac{1}{b^2-a^2}\right)=0
\end{align}

    It is therefore possible to ignore the gravitational contribution of 
cylindrical shells with radii larger than $a$ and consider only the 
internal cylinder of radius $a$. When the previous integral is 
recalculated for the case where $a>b$ it is found that:
\begin{align}
g(a)=\frac{4\pi G\sigma b}{a}
\end{align}

    To transform to a volumetric calculation, $\sigma$ is replaced by a 
three-dimensional mass density $\rho$ and $g$ is summed over cylindrical 
shells from $b=0$ to $b=R_{\rm\,T}$ to find the surface gravity of a torus 
in an infinite cylinder approximation:
\begin{align}
g_{\rm\,T}=\int_0^{R_{\rm\,T}}\frac{4\pi G\rho b}{R_{\rm\,T}}db=2\pi G\rho 
R_{\rm\,T}
\end{align}

\noindent
which also shows that the gravity within an infinite cylinder varies 
linearly with radius. Mass density $\rho$ is assigned to both the toroidal 
and spherical neutron stars. The radius of the neutron sphere is $R_{
\rm\,S}$ and the minor radius of the neutron torus $R_{\rm\,T}$. For a 
sphere, surface gravity $g_{\rm\,S}$ can be immediately calculated:
\begin{align}
g_{\rm\,S}=\frac{GM_{\rm sphere}}{R_{\rm\,S}^2}=\frac{4\pi\,G\rho R_{
\rm\,S}}{3}
\end{align}

    By calculating the pressure at the centre of the sphere ($P_{\rm\,SC}
$) and torus ($P_{\rm \,TC}$), then equating the two values, it will be 
possible to compare the limiting radii at which further gravitational 
collapse takes place. The surface pressures are assumed to be zero and 
integration is performed over infinitesimally thin (spherical or 
cylindrical) shells of matter. The pressure difference between the inner 
and outer surface of a shell is given by the weight of the shell divided 
by the area of the inner shell surface. Noting that the weight of the 
shell depends on the local value of gravity, which is constant throughout 
the shell and a linear function of radius from zero at the centre to $g_{
\rm\,S}$ or $g_{\rm\,T}$ at the surface, one can write:
\begin{align}
P_{\rm\,SC}&=\frac{\rho g_{\rm\,S}}{R_{\rm\,S}}\int_0^{R_{\rm\,S}}
r\,dr=\frac{\rho g_{\rm\,S}R_{\rm\,S}}{2}=\frac{2\pi G\rho^2R_{\rm\,S}^2}{
3}\\
P_{\rm\,TC}&=\frac{\rho g_{\rm\,T}}{R_{\rm\,T}}\int_0^{R_{\rm\,T}}
r\,dr=\frac{\rho g_{\rm\,T}R_{\rm\,T}}{2}=\pi G\rho^2R_{\rm\,T}^2
\end{align}

    Equating $P_{\rm\,SC}$ and $P_{\rm\,TC}$ allows the determination of 
an upper limit for the minor radius of a neutron torus in terms of the 
maximum neutron sphere radius. Note that this result is independent of the 
density of neutron star matter and that the reliability of the result is 
improved by the balancing of the Newtonian approximations:
\begin{align}
R_{\rm\,T_{\rm\,max}}=\sqrt{\frac{2}{3}}\times R_{\rm\,S_{\rm\,max}}
\approx8.5\,{\rm km}
\label{rtorus}
\end{align}

    As might be expected, the minor radius of an infinitely long neutron 
cylinder must be smaller than the maximum spherical radius. In 
circumstances where the infinite cylinder approximation is invalid, the 
minor radius will be further constrained. If general relativity were to be 
used then the pressure gradient for a spherical star would be given by the 
standard equation describing hydrostatic equilibrium:
\begin{align}
\frac{dP}{dr}=-G\frac{(\rho+P/c^2)[m(r)+4\pi r^3P/c^2]}{r[r-2Gm(r)/c^2]}
\end{align}

    Here $P$ is the pressure at some radius $r$ and $m(r)$ is the mass 
enclosed by the 2-sphere defined by $r$, whose internal density may vary 
with radius according to a chosen equation of state. General relativity 
requires larger neutron degeneracy pressures if gravity is to be resisted, 
but the estimate of (\ref{rtorus}) is adequate for the present discussion.

    Confining the discussion to those toroidal stars whose gravitational 
implosion directly results in a TBH rather than intermediate white dwarf 
or neutron density phases, the pre-collapse seed star is assumed to be 
incompressible and of solar density $\sim 1400\,$kg\,m$^{-3}$. For the 
purposes of approximation, the surface areas of extremal Kerr BH event 
horizons are equated with the surface areas of TBHs with equal mass, 
alternatively this may be viewed as equating the entropy of the BHs. The 
TBH is assumed to have an angular momentum equal to the extremal Kerr BH 
of equal mass. In addition, the TBH geometry will be taken to be that of 
an Euclidean torus parameterised by the major and minor radii $R_1$ and 
$R_2$ respectively. This crude model permits the preparation of order of 
magnitude estimates.

    An extremally rotating Kerr BH has $r_+=m$ so its area is $A=4\pi 
r_+^2=4\pi m^2$. The surface area of an Euclidean torus is 
$A=4\pi^2R_1R_2$ so, to a good approximation, the rotating TBH mass is 
related to the TBH area by equating these two expressions for area and, 
after restoring the natural constants ($r_+=Gm/c^2$), it is found:
\begin{align}
M_{\rm TBH}\approx \frac{c^2}{G}\sqrt{\pi R_1R_2}
\label{mtbh}
\end{align}

    Now consider a (low density) toroidal star (TS) of Euclidean geometry 
whose major radius is $R_1$ as before, but with a minor radius $R_3$. 
Evidently $R_3>R_2$ otherwise the TS is a TBH and $R_3<R_1$ ensures the 
star is toroidal. The TS undergoes gravitational collapse once its nuclear 
fuel is exhausted and the resulting TBH is assumed to have the same major 
radius $R_1$ as the TS. Since the volume of the TS is $V_{\rm TS}=2\pi 
R_1R{\,}_3^2$, and the TS is composed of constant density material 
$\rho\sim1400\,$kg\,m$^{-3}$ then the mass of the toroidal star will be 
$M_{\rm TS}=2\pi\rho R_1R{\,}_3^2$. Following a supernova (SN) implosion 
of the star, typically most of the mass will have been ejected. A 
parameter $\eta$ represents the fraction of the original TS mass remaining 
in the TBH after the SN. The remaining mass is identified with the mass of 
the resultant TBH so that:
\begin{align}
\frac{c^2}{G}\sqrt\frac{R_2}{4\pi R_1}\approx\rho\,\eta R_3^2
\label{mrem}
\end{align}

Having already determined the maximum minor radius of a neutron torus in 
(\ref{rtorus}) this implies:
\begin{align}
R_1>R_2\ga8.5\,\text{km}
\end{align}

Taking the limit as $R_3\to R_1$ with $R_3<R_1$ in (\ref{mtbh}) and using 
the relation $M_{\rm TBH}=\eta M_{\rm TS}$ with $\eta=0.1$ (90\% mass 
ejection) gives a limit for TBH formation:
\begin{align}
\frac{R_1^5}{R_2}>\frac{c^4}{4\pi\rho^2\eta^2G^2}\approx7.4\times10^{48}
\,\text{m}^4
\end{align}

Allowing $R_2\to 8.5\,$km with this condition gives a lower bound for 
$R_1$:
\begin{align}
R_1\ga36\times10^9\,\text{m}
\end{align}

    If the TS grows too large and too massive, then it will become a TBH 
without an implosion or electron/neutron degeneracy supported phases. 
Since the area of the TS is larger than the area of a TBH of the same 
mass, then $R_3>R_2$ because $R_1$ is common to both. Consideration of 
(\ref{mrem}) in the case where $R_3\to R_2$ then leads to:
\begin{align}
R_1R_2^3<\frac{c^4}{4\pi\rho^2\eta^2G^2}\approx7.4\times10^{48}\,\text{m}
^4
\end{align}

    The SN is assumed to shed 90\% of the original star's mass during the 
implosion (this assumption is the least reliable and easily dominates the 
combined errors of the remaining assumptions). In special cases where 
$R_2$ approaches $R_1$ and the inequalities hold then almost no mass is 
lost because the star does not collapse much before the event horizon 
engulfs it. Since it has been assumed that the major radius is unchanged 
during collapse,
conservation of angular momentum dictates that the angular velocity of the 
resulting BH will match that of the seed star. Hence, less mass ejection 
is anticipated than in more familiar SN events wherein a star collapses to 
form a spheroidal BH with very high angular velocity. For a given 
$R_2/R_1$ ratio, the permissible range of toroidal star masses which can 
gravitationally collapse to form a TBH range is typically quite broad 
(Fig.~9). This issue is returned to later.

    These massive toroidal stars would have rapidly exhausted their 
nuclear fuel. Regions suitable for fusion reactions occupy a much larger 
proportion of the total stellar volume in toroidal stars than in spherical 
stars. The end result would be a supernova-like implosion, most likely the 
first SN event of its host galaxy, presumably localised to one portion of 
the torus initially, Fig.~2(b). Because the implosion is limited by the 
speed of light, it could take several hours for the implosion to propagate 
around the torus in both directions, Fig.~2(c), until the implosion fronts 
meet at the opposite end of the torus. During the implosion a thin tubular 
event horizon expands along the torus, eventually encountering itself and 
sealing to provide a stable TBH, Fig.~2(d). The illustrations of Fig.~2 
are not based upon precise physical calculations, they are merely intended 
show the progression of the gravitational implosion around the torus.

\begin{figure*}
\centerline{\psfig{figure=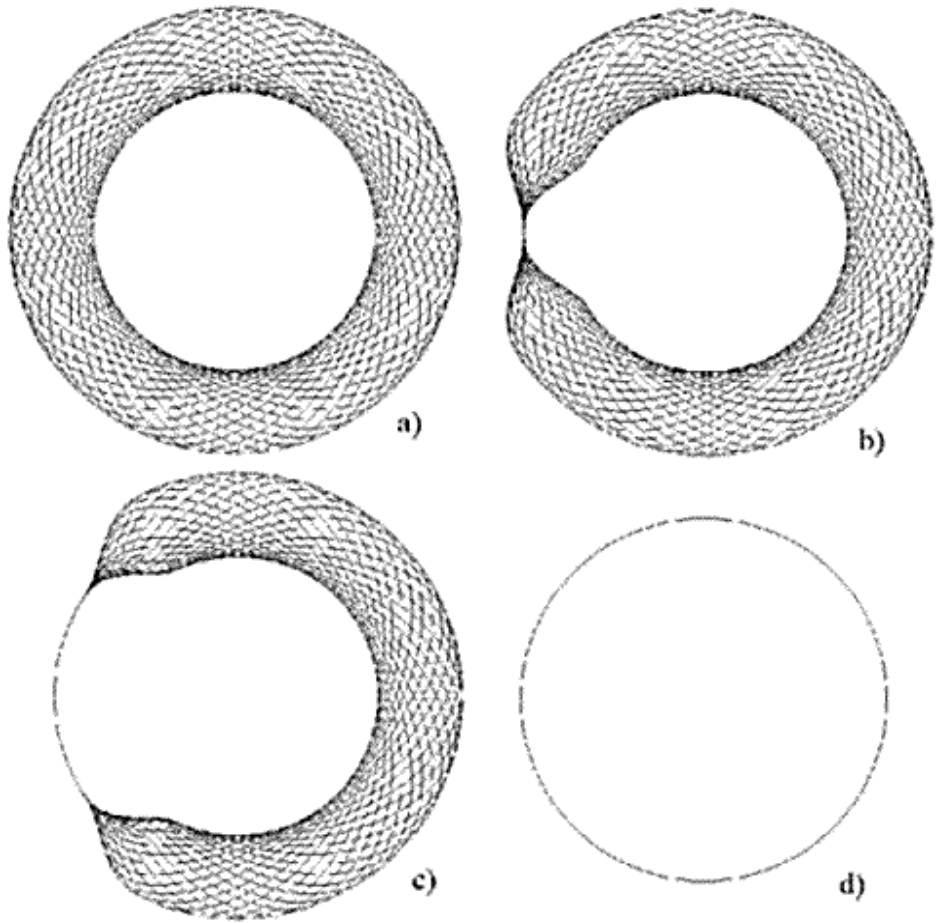,width=0.7\textwidth,angle=0}}
\caption{Gravitational collapse of a toroidal star to a TBH. Collapse 
initially localised to one region of the torus (b) propagates 
bi-directionally around the torus until implosion fronts meet and the TBH 
topology is established.}
\end{figure*}

    The mass of the toroidal star is such that if as much as 90\% of its 
mass is outwardly expelled during the SN implosion, there will still 
remain enough mass to construct what must inevitably become a BH rather 
than a neutron star remnant. Suppose that much more of the mass is ejected 
during the SN, perhaps 99\%, then what may remain could conceivably be a 
toroidal white dwarf or toroidal neutron star. In either case, turbulence 
and dissipative processes are unlikely to leave these delicate structures 
unchanged. If macroscopic axisymmetry is retained, e.g. through 
electromagnetic confinement of the torus, then after a brief period the 
torus will evolve to a smaller major radius. As this occurs, an increase 
in either its minor radius, its density or, more likely both ensues. 
Therefore, toroidal white dwarves could become toroidal neutron stars and 
toroidal neutron stars could become toroidal black holes. Toroidal neutron 
stars of masses $\sim10^6\msun$ are precluded as serious AGN candidates by 
their limited lifespan, slender geometry and inability to endure sustained 
accretion. Smith and Mann (1997) have recently investigated gravitational 
collapse as a TBH formation mechanism starting with collisionless 
particles of random velocities but zero net angular momentum.

    Quasar observations yield spectra with very strong metallic absorption 
lines. The population II stars of the galactic centre would mainly consist 
of Hydrogen and Helium, which has previously troubled spheroidal BH quasar 
models. Because SNe are efficient at generating heavy elements, the TBH 
creation SN would have scattered a substantial amount of metallic elements 
into the ambient galactic environment, imparting its signature on the 
radiation spectrum of the central engine.

\section{STABILITY OF ROTATING TOROIDAL BLACK HOLES}

    For some time, following the work of Hawking (1972) and Hawking \& 
Ellis (1973), it was thought that TBHs were unstable, albeit marginally. 
This somewhat contra-intuitive result assumed that Einstein's cosmological 
constant ($\Lambda$) was zero. Numerical computations of collisionless 
particles resulting in a transient toroidal event horizon (terminating in 
a sub-extremal Kerr BH) and assuming $\Lambda=0$ were performed by 
Abrahams et al (1994), Hughes et al (1994) and Shapiro, Teukolsky \& 
Winicour (1995). These results were consistent with the topological 
censorship theorem of Friedman, Schleich \& Witt (1993) which implies that 
a light ray cannot pass through the central toroidal aperture before the 
topology becomes spherical. More recently, papers by Huang \& Liang 
(1995), Aminneborg et al (1996), Mann (1997), Vanzo (1997) and Brill 
(1997) have provided mathematical descriptions of TBHs within the 
framework of general relativity. These equations assume that the 
cosmological constant is negatively valued to admit stability for the TBH 
and is literally constant throughout the spacetime described, which has an 
anti-de Sitter (AdS) background. The Vanzo paper claims that a TBH can 
exist in a virtually flat spacetime because the TBH size is determined by 
the mass and conformal class of the torus, not by the cosmological 
constant. Rotating charged black (cosmic) strings have been described by 
Lemos \& Zanchin (1996). A spacetime metric for a rotating, uncharged TBH 
presented by Klemm, Moretti \& Vanzo (1998), is hereafter referred to as 
the KMV metric. This metric is not unique, but it is the first 
generalisation to admit rotation of TBHs. Holst \& Peldan (1997) showed 
that rotating Banados-Teitelboim-Zanelli (BTZ) BHs cannot be described in 
terms of a 3+1 split of spacetime, instead spacetimes of non-constant 
curvature are required.

    Physical measurements to date have been unable to establish 
conclusively whether $\Lambda$ is positive or negative. The accelerating 
cosmological recession of distant SNe favours a positive value, though 
whether this recession is attributable to a cosmological constant is the 
subject of continuing debate. Arguments against TBH stability have assumed 
that in our universe, the constant is precisely zero everywhere. The weak 
energy condition is assumed to be satisfied, although it is known to be 
violated in certain situations e.g. Casimir effect and Hawking radiation. 
Topological BHs in anti de-Sitter spacetimes are now known not to conflict 
with the Principle of Topological censorship, for a recent discussion, see 
Galloway et al (1999). Intuitively, rotating TBHs are not dissimilar to 
Kerr BHs in that both contain ring singularities whose radii are 
determined by the angular momentum assuming constant BH mass. One extra 
parameter is necessary to characterise a stationary TBH in addition to the 
mass, angular momentum and charge of the Kerr-Newman metric. This 
parameter determines the exact geometry of the torus and can be expressed 
as the ratio of the minor and major radii $R_2/R_1$ (as used here) or the 
ratio akin to a Teichm\"uller parameter presented by the KMV paper. 
Stationarity is preserved only when this parameter achieves a balance with 
the TBH mass and angular momentum, and to a lesser degree the charge.

    It was demonstrated by Gannon (1976) that for non-stationary BHs in 
asymptotically flat spacetimes, the topology of the event horizon must be 
either spherical or toroidal. A rotating TBH located at the centre of a 
galaxy surrounded by accreting matter is manifestly non-stationary. The 
stationary BH metrics containing physical singularities are acutely 
idealised, the Kerr metric contains a ring singularity surrounded by 
vacuum i.e. a universe devoid of other matter. This is a gross 
simplification of what would be found in nature. Inside the inner event 
horizon $r_-$, particles are not compelled to collapse towards the 
singularity, but are free to explore all radii $0\leq r\leq r_-$. Suppose 
a Kerr BH forms by the collapse of a non-rotating neutron star. The matter 
at the surface of the neutron star can reach the singularity in a finite 
proper time. On the other hand, viewed from infinity, this matter never 
crosses the event horizon, less still reaches the singularity. According 
to distant observers, the matter is frozen fractionally above the event 
horizon. Just as infalling matter experiences the crossing of outer then 
inner event horizons in finite time, it also witnesses the end of the 
external universe before nearing the singularity. The only possible answer 
to the question: ``when does the BH become stationary to distant 
observers?'' is never. Indeed, one might venture that truly stationary 
spacetimes are forbidden. It is dangerous to be guided by predictions 
about BH stability which rely on stationarity as one of the underlying 
assumptions.

    Perhaps there is some deeper significance underlying the 
unobtainability of stationarity. Consider a closed universe approaching a 
big crunch and contracting rapidly in all directions. The surface defining 
the outer reaches of this universe could be considered as the event 
horizon of a BH beyond which spacetime does not exist in the usual sense. 
This is a BH that could conceivably approach stationarity in a short and 
finite time as measured by the clocks of all internal observers, there 
being no external observers. A singularity develops which is accessible to 
all the infalling matter. The outermost layers of the imploding universe 
catch up with the innermost layers at the Cauchy horizon, the surface of 
infinite blue shift. A vacuum develops in the region surrounding the 
singularity as it swiftly becomes devoid of matter and stationarity is 
achieved. The singularity now contains the entire mass of the pre-collapse 
universe and the Pauli exclusion principle does not participate in the 
physics of the singularity. What grounds are there for discarding the 
Pauli exclusion principle? This principle has successfully predicted the 
existence of white dwarfs and neutron stars. Could quark degeneracy arise? 
What might string theory predict? There are obvious similarities between 
the Pauli exclusion principle and the premise that stationary BHs are 
forbidden. If it is true that BHs truly abhor stationarity then presumably 
re-expansion would be the only option.

    Suppose that a TBH with a substantial central aperture is rotating in 
asymptotically flat space with a near maximal angular momentum (event 
horizon velocity approaching the speed of light). In principle, there is 
no reason why the rotational energy of this TBH cannot be arbitrarily 
larger than the TBH's rest mass, whereas a Kerr BH can only hold at most 
29\% of its total energy in rotational form, the remainder being the 
irreducible mass. According to topological censorship, the TBH must become 
spheroidal before a light ray can traverse the aperture. The fate of the 
excess rotational energy is something of a conundrum. Is the excess energy 
hastily expelled by some undiscovered mechanism?  Is topological 
censorship flawed? Would the TBH break up into multiple co-rotating 
spheroidal BHs? Does the Kerr BH rotate above the extremal limit, and if 
so is the singularity revealed? These problems can be circumvented for now 
by assuming a negative $\Lambda$.

    It seems somewhat coincidental that the cosmological constant is so 
nearly zero and not very much larger in value, on purely theoretical 
grounds a value 120 orders of magnitude greater than observational limits 
might have been expected. One plausible suggestion was proposed by Coleman 
(1988). According to the author, macroscopic cancellation mechanisms 
operate on the zero point energies under normal circumstances and these 
result in a zero expectation for $\Lambda$. The situation is, however, 
complicated in the presence of intense gravitational fields generated by 
BHs, particularly in the immediate vicinity of the singularities residing 
within the event horizons. Under such conditions, the cancellation of zero 
point energies operates imperfectly and gives rise to what may be 
considered a localised but substantial cosmological `constant'. By this 
means, TBH stability could be ensured within a universe where elsewhere 
$\Lambda$ is small. It was also suspected that zero point energy might 
play a part in the physics of curved spacetimes because of imperfect 
cancellations complicating the assumptions underlying the quantum 
mechanical technique of renormalization (Misner, Thorne \& Wheeler 1973). 
It comes as little surprise that quantum effects may play a prominent role 
when the spacetime of classical general relativity becomes singular, the 
stability of the TBH structure could prove to be the only direct evidence 
of this.

    The KMV metric has axial symmetry and the horizons are Riemannian 
surfaces of constant gravity that obey the familiar BH entropy-area laws. 
Utilising the membrane paradigm approach (Thorne, Price \& MacDonald 1986) 
simplifies the consideration of the physics of these BHs outside the event 
horizon. The fact that these objects are thermodynamically well behaved, 
whilst interesting, is of little relevance to the present discussion. 
Parallels between rotating TBH solutions and the Kerr solutions for 
spinning spheroidal BHs may be drawn, for instance both have ergoregions 
external to their event horizons and the maximum angular momentum of each 
is bounded for a given mass. Conservation of mass and angular momentum is 
known to be satisfied. A maximally rotating Kerr BH has a static limit 
extending to $2m$, double the radius of the outer event horizon. The 
equator of the static limit surface circles with the speed of light at 
extremality. The ergoregion occupies the region between the static limit 
and the event horizon within which everything is compelled to co-rotate 
with the BH due to the spacelike character of the time coordinate. 
Similarly, the maximal KMV metric determines the ratio $r_{\rm s}/r_+$ to 
be 1.59. Orbits within 300\% of the extremal Kerr event horizon radius are 
unstable and matter (the accretion disk) tends to be drawn towards the BH. 
For a maximally rotating BH, entering the ergoregion becomes impossible 
because incoming particles would have to travel `faster' than light and 
possess an infinite amount of energy. Similarly, if the BH is rotating 
slightly below this rate then only a tiny fraction of the external 
particles will penetrate the ergoregion, those with very high kinetic 
energies.

\begin{figure*}
\centerline{\psfig{figure=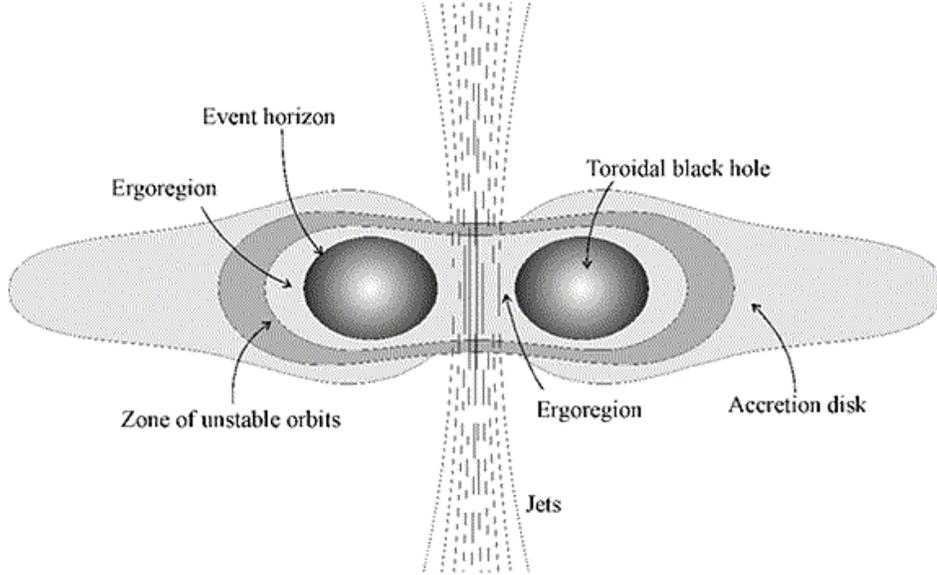,width=0.7\textwidth,angle=0}}
\caption{Meridional section of a rotating TBH producing jets when matter 
accretes from a
surrounding disk illustrating the approximate form of the ergoregion and 
limiting boundary
for stable orbits.}
\end{figure*}

    An ergoregion enshrouds the toroidal event horizon of the KMV metric. 
Fig.~3 depicts a cross-sectional view of a rotating TBH. The event horizon 
will be enshrouded by an ergoregion which, depending upon the precise 
geometry of the TBH, might entirely seal the central aperture. Beyond the 
ergoregion lies what is sometimes referred to as a zone of unstable orbits 
within which particles are unable to establish repeating orbital patterns 
by following geodesic pathways. The ergoregion does not intersect the 
event horizon at any point, as it does at the poles of a Kerr BH. 
Particles cannot penetrate the ergoregion of a maximally rotating TBH, 
whichever trajectory is attempted. The maximal rotation rate will not be 
achieved in practice because the BH is able to reduce its rotation rate by 
several methods which are relevant to jet formation and several 
theoretical reasons such as the fact that the internal singularity would 
become naked, even as viewed from infinity.

\section{METRIC OF ROTATING TOROIDAL BLACK HOLE}

    The KMV metric of an uncharged rotating TBH in asymptotically anti 
de-Sitter (AdS) spacetime  is tentatively forwarded as a model for the 
naturally occurring TBH. The primary reservations concerning the physical 
applicability of the topological BH metrics in AdS gravity are that 
$\Lambda$ is assumed to be independent of location and, contrary to the 
most reliable observations, negative in value. Given the uncertainty 
regarding the role of quantum mechanics in BH physics, these assumptions 
may be invalid. The metric describes a vacuum solution of Einstein's 
equation which has reached equilibrium after an infinite coordinate time 
has elapsed. It possesses a ring singularity, but no provision for 
accreting matter has been made. Indeed, a massive accretion disk may act 
as a stabilising influence on a TBH within an asymptotically flat 
spacetime (further discussed in appendix~B). A negative cosmological 
constant may be thought of as contributing a cosmological attraction. In 
its absence, the combination of a host galaxy's matter and the nearby 
massive accretion disk surrounding the outer periphery of a TBH located 
within a galactic nucleus may provide a natural substitute for the 
stabilising negative $\Lambda$ used in the AdS metrics.

    With these considerations in mind, attention is focused on the KMV 
metric which, for convenience, is now recalled:

\begin{align}
ds^2=-N^2dt^2+\frac{\rho^2}{\Delta_r}dr^2+\frac{\rho^2}{\Delta_P}
dP^2+\frac{\Sigma^2}{\rho^2}(d\phi-\omega\,dt)^2
\label{metr}
\end{align}
where the following relations apply:
\begin{align}
\rho^2&=r^2+a^2P^2\\
\Sigma^2&=r^4\Delta_P-a^2P^4\Delta_r\\
\label{delr}\Delta_r&=a^2-2mr+r^4/l^2\\
\Delta_P&=1+\frac{a^2P^4}{l^2}\\
N^2&=\frac{\rho^2\Delta_P\Delta_r}{\Sigma^2}\\
\omega&=\frac{\Delta_rP^2+r^2\Delta_P}{\Sigma^2}a
\end{align}

    Here, the angular velocity is $\omega$, the equatorial angle is 
$\phi$, $P$ is another angular variable with some period $T$, $r$ is a 
pseudo-radial coordinate, $a$ is the angular momentum per unit mass and 
$l$ is defined as $\sqrt{-3/\Lambda}$. The ratio of $T$ to $2\pi$ is 
analogous to the Teichm\"uller parameter describing a flat torus in 
Riemannian geometry. TBH mass by the ADM definition is $M=mT/2\pi$ and 
angular momentum $J=Ma$. The coordinate $r$, as in the  Boyer-Lindquist 
form of the Kerr metric, is only a true radial coordinate as $r\to\infty$ 
with $r=0$ the location of a ring singularity not corresponding to zero 
radius. Unlike the Kerr BH, as $r\to-\infty$ within the equatorial plane, 
this point is outside the event horizon. It would be preferable to 
introduce a coordinate transformation whereby $r^\prime=f(r)$ such that 
$f(\infty)=\infty$, $f(0)=a$ (say) and $f(-\infty)=0$ and select $f(r)$ 
such that $r^\prime$ is an affine parameter, but this is beyond the scope 
of the present discussion.

    In order for the metric to describe a torus, $P$ is a periodic 
variable with period $T$ and is covered by four patches 
$P=\lambda\sin\vartheta$ at $\vartheta=0$ and $\vartheta=\pi$, and 
$P=\lambda\cos\vartheta$ at $\vartheta=\pi/2$ and $\vartheta=3\pi/2$ where 
$\lambda$ is a constant such that $T=2\pi\lambda$. Between these points 
the behaviour is defined by $\cos\vartheta$ being some $C^\infty$ function 
(infinitely differentiable) of $\sin\vartheta$ and vice versa.

    Upon inspection of the metric, (\ref{metr}), it can be seen that 
$\Delta_r$ in (\ref{delr}) becomes zero at the event horizon. Inner and 
outer event horizons exist as real and positive roots of the quartic 
equation with real coefficients: $r^4-2ml^2r+a^2l^2=0$ along with two 
other physically less meaningful complex conjugate roots. For the 
extremally rotating case, $a=a_c=\sqrt{3}\times\root3\of{lm^2/4}$ and 
these two roots coincide. It is straightforward to verify that the real 
roots are $r_\pm=\root3\of{ml^2/2}$.

    The ergoregion is defined as the region between the outer event 
horizon and the static limit hypersurface at which the metric coefficient 
of $dt^2$ vanishes altogether, i.e. $g^{\,tt}=\omega^2\Sigma^2/\rho^2-N^2$ 
which is solved for $r$ by another quartic $r^4-2ml^2r-a^4P^4=0$. This 
polynomial has one real and positive root, one real and negative root and 
two complex conjugate roots. The real positive root has a minimum value of 
$\root3\of{ml^2}$ for $P=0$ which is larger than $r_+$ by a factor of 
1.59. The static limit hypersurface is well separated from the event 
horizon and, unlike the poles of the Kerr situation, these surfaces are 
nowhere contiguous. Therefore, a substantial ergoregion is observed.

Examining the equatorial plane by setting $dP=P=0$ the metric reduces to:
\begin{align}
ds^2=-\left(\frac{r^2}{l^2}-\frac{2m}{r}\right)dt^2+\left(\frac{r^2}{l^2}
-\frac{2m}{r}+
\frac{a^2}{r^2}\right)^{\!-1}\!\!\!dr^2
\label{eqmet}
\end{align}

    In order to determine the trajectories of null geodesics within this 
hypersurface use can
be made of the Euler-Lagrange equations with $K=ds^2/2\,$:
\begin{align}
\frac{\partial K}{\partial x^a}-\frac{d}{du}\left(\frac{\partial K}{
\partial\dot x^a}
\right)=0
\end{align}
the overdot denoting differentiation with respect to some affine parameter 
$u$. The equatorial metric (\ref{eqmet}) may be partial differentiated 
with respect to $t$ and $\phi$ respectively then integrated with respect 
to $u$ to give:
\begin{gather}
\left(\frac{2m}{r}-\frac{r^2}{l^2}\right)\dot t-a\dot\phi=\balpha\\
r^2\dot\phi-a\dot t=\bbeta
\end{gather}
Where $\balpha$ and $\bbeta$ are constants of integration. A third 
constant $\bgamma=\balpha/\bbeta$ can be defined and used to relate both 
equations:
\begin{align}
\left(\frac{2m}{r}-\frac{r^2}{l^2}\right)\dot 
t-a\dot\phi=\bgamma(r^2\dot\phi-a\dot t)
\end{align}
which upon rearrangement reads:
\begin{align}
\frac{d\phi}{dt}=\frac{\dot\phi}{\dot t}=\frac{2ml^2-r^3+\bgamma arl^2}{
arl^2+\bgamma l^2r^3}
\end{align}

    Next, boundary conditions are imposed by considering the extremal case 
$a=a_c$ for which the angular velocity at the event horizon $r_+$ is, 
using the expressions for $a_c$ and $r_+$ and noting that $\omega=a/r^2$ 
in the equatorial plane.
\begin{align}
\frac{d\phi}{dt}\Bigl |_{r=r_+}=\Omega_H=\frac{a_c}{r_+^2}=\frac{\sqrt3}{l}
=\frac{2ml^2-r_+^3+\bgamma ar_+l^2}{ar_+l^2+\bgamma l^2r_+^3}
\end{align}

Some algebra reveals that the constant of proportionality $\bgamma$ obeys 
the relations:
\begin{align}
\bgamma=\frac{3\root3\of2ml-2\sqrt3a\root3\of{ml^2}}{
\sqrt3\root3\of2ml^2+2al\root3\of{ml^2}}=\frac{3r_+^2-\sqrt3al}{
\sqrt3lr_+^2+al^2}
\end{align}

So now the rate of change of $\phi$ with respect to coordinate time $t$ is 
fully determined. Following straight from the metric and the condition 
that $ds=0$ for null geodesics:
\begin{align}
\left(\frac{2m}{r}-\frac{r^2}{l^2}\right)\!\!\dot t^2\!\!+\!\!\left(\frac{
r^2}{l^2}-\frac{2m}{r}+\frac{a^2}{r^2}\right)^{\!\!-1}\!\!\!\!\!\!\dot 
r^2\!\!\!+\!r^2\!\dot\phi^2\!\!\!-\!2\,a\,\dot\phi\,\dot t=0
\end{align}

Dividing throughout by $(dt/du)^2$ eliminates the affine variable allowing 
$dr/dt$ to be found using the previously derived expression for 
$d\phi/dt$:
\begin{align}
\frac{dr}{dt}=\!\sqrt{\left[2a\frac{d\phi}{dt}-r^2\frac{d\phi}{dt}
^2\!\!+\!\frac{r^2}{l^2}-\frac{2m}{r}\right]\!\!\left[\frac{r^2}{l^2}
-\frac{2m}{r}-\frac{a^2}{ r^2}\right]}
\end{align}

    It would be possible to continue this analysis by integrating with 
respect to $t$ for each variable $r$ and $\phi$, resulting in cumbersome 
mathematical terms. It is sufficient for now to say that these equations 
allow the null congruences of the equatorial plane to be readily 
determined by numerical methods.

\section{ROTATING TOROIDAL BLACK HOLE IN ASYMPTOTICALLY FLAT SPACE}

    In order to visualise a TBH in asymptotically flat space and its 
effect on local spacetime, a method which approximates the time dilation 
at locations in space surrounding arbitrarily complex mass configurations 
is now introduced. First, the time dilation is derived for the 
Schwarzschild spacetime with metric:
\begin{align}
ds^2\!\!=\!\!\left(\!1\!-\!\frac{2m}{r}
\!\right)\!dt^2\!\!-\!\left(\!1\!-\!\frac{2m}{r}\!\right)^{\!\!-1}
\!\!\!\!\!dr^2\!\!-\!r^2(d\vartheta^2\!\!+\sin^2\!\!\vartheta 
\,d\vartheta^2)
\end{align}

    The event horizon of this static spacetime occurs when $g^{rr}$ 
becomes infinite, or $r_+=2m$ in geometrical units. Consider the time 
dilation of a stationary particle located at some constant $r$, 
$\vartheta$ and $\phi$. The metric interval $ds$ can be interpreted as the 
proper time of particles travelling on timelike paths so that $d\tau=ds$. 
The time dilation may be read from the metric at once as:
\begin{gather}
\label{tdil}
\frac{d\tau}{dt}=\sqrt{\left(1-\frac{2m}{r}\right)}=\sqrt{\left(1-\frac{
r_+}{r}\right)}=
\sqrt\psi\\\text{where\quad}\psi=\left(1-\frac{2m}{r}\right)
\end{gather}

    As $r\to\infty$ notice that $d\tau/dt=1$ whilst $d\tau/dt$ decreases 
towards zero as the event horizon is approached, as expected. Now, the 
particle is allowed to undergo radial motion $dr\neq0,d\vartheta=d\phi=0$. 
The metric is divided throughout by $dt^2$ and the particle's radial 
velocity in local coordinates is given by $v_p=dr/d\tau$. This yields a 
similar equation to the last but with the introduction of a $v_p$ 
dependent term:
\begin{align}
\frac{d\tau}{dt}=\frac{\psi}{\sqrt{\psi+v_p^2}}
\end{align}

    For the Schwarzschild BH, the event horizon and stationary limit 
coincide at $r=r_+$ and $d\tau/dt$ becomes zero there. A radial velocity 
can affect the time dilation but cannot alter the location of the 
hypersurface at which the time dilation approaches zero. Conversely, in 
the Kerr case which is now briefly addressed, $d\tau/dt$ becomes zero for 
stationary particles outside the event horizon on the stationary limit, 
the outermost boundary of the ergosphere. Particles motionless with 
respect to distant observers will appear to freeze at the static limit but 
particles in prograde orbits can both penetrate and escape the ergosphere 
in a finite coordinate time. For retrograde orbits, the time dilation 
approaches zero at radii beyond the static limit so the location of the 
stationary limit is meaningful only for particles with zero coordinate 
velocity. The ergosphere is a zone where some particles are able to travel 
on spacelike trajectories --- these trajectories becoming increasingly 
probable close to the outer event horizon. Whereas negative energy states 
are only available within the ergosphere of a Kerr BH, a charged 
Kerr-Newman BH offers negative energy states beyond the static limit.

    Returning to the Schwarzschild metric, situations where the particle 
undergoes transverse (azimuthal) motion are examined by setting $dr=0$ and 
$d\vartheta=0$. Noting that $v_p=r^2d\phi/d\tau$ this leads to:
\begin{align}
\frac{d\tau}{dt}=\sqrt\frac{\psi}{1+v_p^2}
\end{align}

    The next task in this analysis is to derive approximations for the 
time dilation experienced by observers nearby a moving point mass where 
the clock at infinity is motionless relative to the nearby observers. This 
cannot be read directly from the Schwarzschild coordinates since the mass 
of the singularity is stationary with respect to observers at infinity. By 
taking the limit as $m\to0$ one obtains $\psi\to1$ and both the previous 
equations reduce to the time dilation of special relativity when two 
objects are in relative motion. These limits are used to introduce a 
contribution to the time dilation equivalent to inducing a motion of the 
clocks at infinity. The situation then describes clocks at infinity moving 
with velocity $v_p$, clocks of local observers moving with velocity $v_p$ 
and a motionless point mass. Since all inertial frames are equivalent, one 
can think of this as stationary clocks and a moving mass with velocity 
$v_p$ in the opposite direction. The various possibilities are depicted in 
Fig.~4. Condition 6 has been determined by taking the ratio of the 
expression in condition 4 with the expression in condition 2. Likewise, 
condition 7 has been determined by taking the ratio of the expression in 
condition 5 with the expression in condition 3. By taking these ratios, 
Lorentzian boosts are applied which remove the time dilation contributions 
of expressions 4 and 5 which were purely due to the relative motions of 
the clocks. What remains are motionless clocks in the presence of a moving 
point mass.

\begin{figure*}
\centerline{\psfig{figure=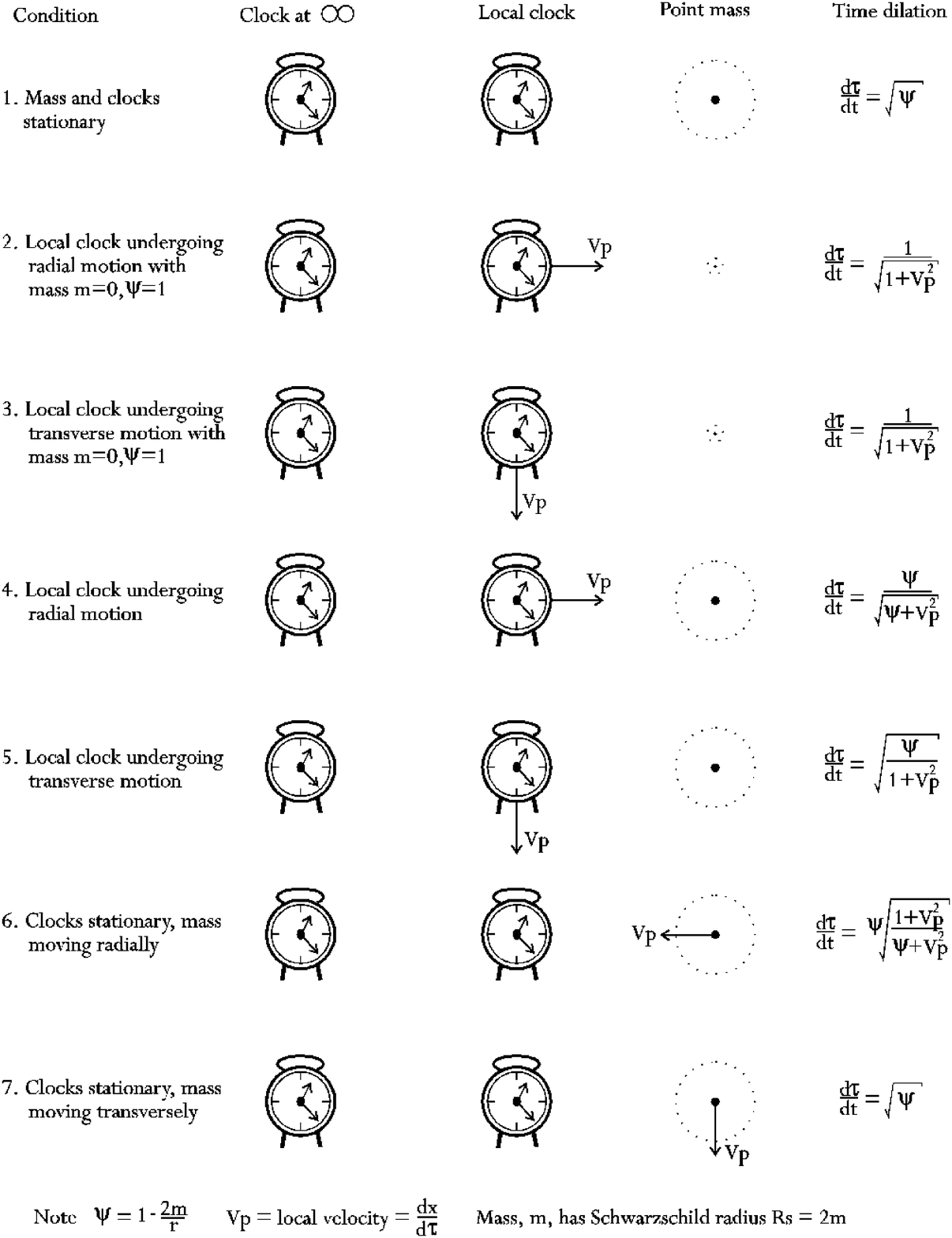,width=0.8\textwidth,angle=0}}
\caption{Relative time dilation in Schwarzschild gravity for local 
($2m<r<\infty$) and remote
($r\to\infty$) clocks under various limiting circumstances. Time dilation 
for
stationary clocks in the presence of a moving mass depend on the velocity 
and direction of
the black hole's motion.}
\end{figure*}

    The expressions in conditions 2 and 3 are identical implying that time 
dilation between observers in the absence of gravity is independent of the 
direction of motion. Conditions 2 and 3 are limiting cases of conditions 4 
and 5 respectively in the absence of matter. The parity between the 
expressions of conditions 1 and 7 suggests that only the component of the 
mass's velocity towards the local clock (not the clock at $\infty$ since 
this is always unaffected by the mass) contributes to the time dilation of 
the local clock relative to the clock at infinity. Condition 6 can then be 
used to calculate the time dilation precisely in more general 
circumstances providing that $v_p$ is the velocity component towards the 
local clock. Note also that there is no requirement for the clocks and the 
mass to be aligned as they are in Fig.~4, the expressions presented are 
valid for all configurations owing to the perfect spherical symmetry of 
the Schwarzschild geometry.

    Suppose the Schwarzschild point singularity is subdivided into $N$ 
smaller but not necessarily equal masses, each point mass being located at 
$r=0$, the same spatial position as the parent singularity. In order to 
accurately recover the time dilation of (\ref{tdil}), one is obliged to 
perform $N$ summations of the ratios $r_+/r$ where $r_+$ relates to the 
Schwarzschild radius of the mass of each child singularity in turn 
according to the equation $r_+=2m_{\rm child}$. This will be generalised 
for the purposes of approximation such that the point masses are not 
coincident but are located separately in space. Thus the distance $r$ will 
in general be different for each point mass. Restoring natural constants, 
the following equation is obtained:
\begin{align}
\frac{d\tau}{dt}=\sqrt {1-\frac{2G}{c^2}\times\sum_n\frac{M_n}{R_n}}
\end{align}

    This may be thought of as a pseudo-principle of superposition and 
these results may be used to approximate an asymptotically flat spacetime 
containing a ring singularity. Firstly, the discussion is confined to a 
momentarily stationary ring singularity i.e. one with zero angular 
velocity and a radius $R_1$ whose derivative with respect to time is 
momentarily zero. An expression for the time dilation relative to 
observers at infinity experienced by a spatially fixed observer due to the 
momentarily motionless ring singularity is derived. The singularity is 
assigned a constant mass per unit length $b$ and radius $a$ such that the 
total mass is $2\pi ab$. The time dilation within the plane of the ring is 
first considered. By symmetry, the only independent coordinate is the 
radius $r$ and the time dilation $d\tau/dt$ at that point is approximated 
by:
\begin{align}
\frac{d\tau}{dt}=\sqrt{1-\frac{2abG}{c^2}\int_0^{2\pi}\frac{d\phi}{\sqrt{
r^2+a^2+2ar\cos \phi}}}
\end{align}

Setting $\varphi=\phi/2$ the time dilation can be expressed in terms of a 
complete elliptic integral of the first kind, K(k):
\begin{gather}
\frac{d\tau}{dt}=\sqrt{1-\frac{8abG}{(a+r)c^2}\int_0^{\pi /2}\frac{
d\varphi}{\sqrt{1-k^2\sin^2\varphi}}}\\
{\text or\quad}\frac{d\tau}{dt}=\sqrt{1-\frac{8abG}{(a+r)c^2}K(k)}
\text{\quad where\quad }k=\frac{2\sqrt{ar}}{a+r}
\label{kpar}
\end{gather}

The point at the centre of the ring ($r=0$) is a special case which is 
readily integrated to give:
\begin{align}
\frac{d\tau}{dt}=\sqrt{1-\frac{4\pi bG}{c^2}}
\label{cring}
\end{align}

    In order to describe a TBH, the ring density $b$ must be smaller than 
$c^2/4\pi G$. By writing $(\sqrt a-\sqrt r)^2\geq0$ and expanding it is 
found that $k\leq1$ in all cases of (\ref{kpar}) satisfying the 
requirements of the elliptic integral, $k$ being the ratio of the 
geometric and arithmetic means of the parameters $r$ and $a$. For the 
static case, the term within the square root becomes zero at the event 
horizon. If rotation is allowed, the time dilation will become infinite 
not at the event horizon but at the static limit, the external boundary of 
the ergoregion where the invariant interval of motionless particles is 
lightlike.

    When the ring singularity rotates with constant angular velocity 
$\omega$, the velocity of a point on the ring is taken to be 
$v_r=a\omega$. Consider the component of this velocity directed towards 
the observer $P$ situated at some radius $r$ from the centre of the ring 
and within the equatorial plane. This component contributes to the time 
dilation experienced by the observer according to the expression presented 
in condition 6 of Fig.~4. The terms within the square root causing a 
deviation from parity of proper and coordinate time are once more summed. 
The ring's total mass $2\pi ab$ as before. Assuming the centre of mass to 
be located at the centre of symmetry, the following estimate of the ring's 
angular velocity shall be used: $a^2\omega^2\approx2\pi bG$ so that the 
angular velocity is $\omega\approx\sqrt{2\pi bG}/a$. Recalling the 
expression for time dilation of condition 6 and substituting 
$\psi=1-(2m/r)=1-r_+/r$ then rewriting in such a way as to give a separate 
and integrable deviation from unity within the square root gives:
\begin{align}
\frac{d\tau}{dt}=\psi\sqrt\frac{1\!+\!v_p^2}{\psi\!+\!v_p^2}=\!\sqrt{
1\!-\!\frac{r_+}{r}\!
\left[\frac{(1\!+\!2v_p^2)-\frac{r_+}{r}(1\!+\!v_p^2)}{1+v_p^2-(\frac{r_+}{
r})}\right]}
\end{align}

    As $r\to r_+$, $d\tau/dt\to0$ which means only particles travelling at 
the speed of light can remain on the horizon, as expected. As it stands, 
this formula allows the deviation from unity within the square root to be 
summed for an arbitrarily large number of point masses, regardless of the 
mass contained by each. Simplification is possible if it is assumed that 
all these point masses are infinitesimally small so that the Schwarzschild 
radius of each is negligible compared to the distance between each mass 
and the local clock where $d\tau/dt$ is to be determined, $r_+\ll r$. 
Implementing this simplification and including the integration symbol to 
emphasise the fact that the point masses should be vanishingly small 
yields:
\begin{align}
\frac{d\tau}{dt}=\sqrt{1-\int\frac{r_+}{r}\left(\frac{1+2v_p^2}{1+v_p^2}
\right)}
\label{vdil}
\end{align}

    The time dilation relative to observers at spatial infinity is now 
derived for points surrounding the rotating ring singularity. These test 
points are assigned cylindrical coordinates $(r,\phi,z)$, and are not 
confined to the equatorial plane. By symmetry the $\phi$ coordinate is 
redundant. The resulting time dilation resembles the previously derived 
expression containing an elliptic integral but with additional complexity:
\begin{align}
\frac{d\tau}{dt}=\sqrt{1-2ab\int_0^{2\pi}\!\!I(\phi)d\phi}
\end{align}
substituting $u=a^2+r^2+z^2$ the integrand reads:
\begin{align}
I(\phi)\!=\!\frac{u+2ar\cos\phi+2a^2r^2\omega^2\sin^2\phi}{\sqrt{
u\!+\!2ar\cos\phi}
\left(u\!+\!2ar\cos\phi\!+\!a^2r^2\omega^2\sin^2\phi\right)}
\label{idil}
\end{align}

    Note that the time dilation at the centre of the ring ($r=0,z=0$) is 
still given by (\ref{cring}) because the velocity of each point mass is 
perpendicular to the line connecting the point mass to the local observer 
and that this holds along the entire axis of rotation. It would be 
possible, but more complicated, to determine the approximate location of 
event horizons using a similar method. One would need to transform the 
local observers to those of a locally non rotating frame (LNRF). This 
would be achieved in the equatorial plane by relating the angular velocity 
of the ring to that of the local observers. Starting with condition 6 of 
Fig.~4, one would generalise to the case where the local clock and point 
mass have separate (non-zero) velocities with respect to the clock at 
spatial infinity by applying a Lorentzian boost to the coordinate clock. 
Then, equivalents to the expression in (\ref{cring}) and (\ref{vdil}) 
would need to be found.

\begin{figure*}
\centerline{\psfig{figure=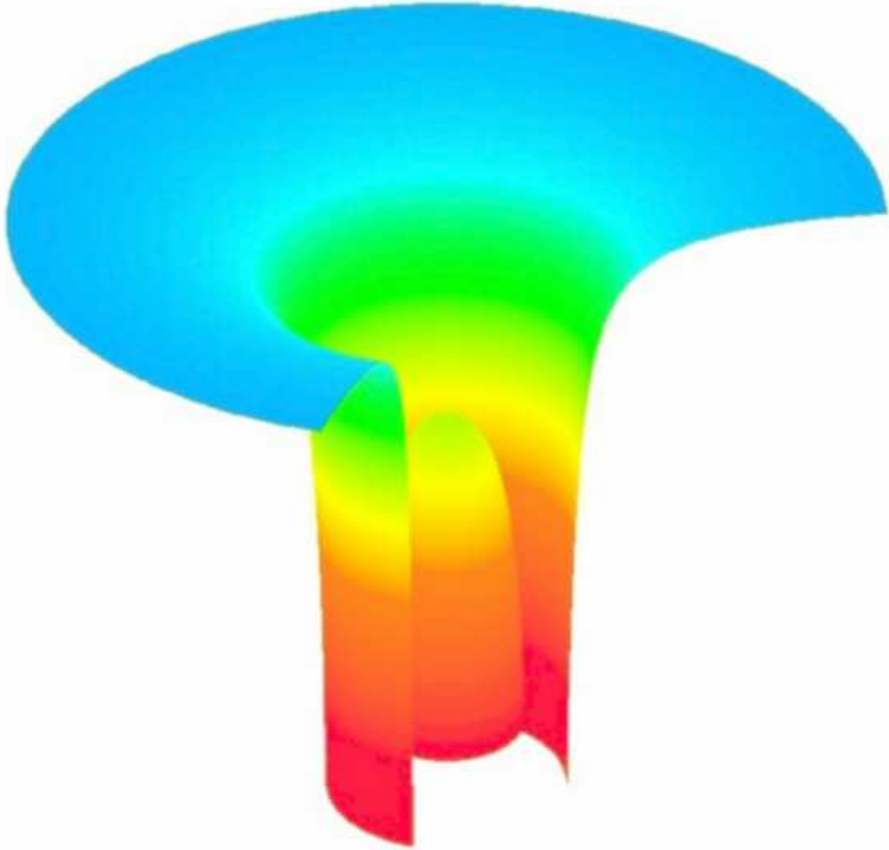,width=0.7\textwidth,angle=0}}
\caption{Embedding diagram for the simulated time dilation within the 
equatorial plane of a rotating TBH. A portion of the outer funnel has been 
omitted revealing the central aperture. The rotation rate in equilibrium 
cannot be precisely estimated, but for any given toroidal geometry, 
relativistic rotation acts to enhance the time dilation at all locations 
except the axis of rotation axis.}
\end{figure*}

    The time dilation can be computed numerically but care is needed when 
selecting the ring's angular velocity $\omega$ otherwise the situation 
becomes unphysical with frame-dragging velocities in excess of $c$. This 
formula was used to determine the shape of the ergoregion in Fig.~3 when 
viewed in cross section. Fig.~5 presents a 3-dimensional projection of the 
time dilation as viewed by observers located at spatial infinity for the 
equatorial plane intersecting a rotating TBH. This embedding diagram 
portrays local time dilation (as viewed by distant observers) due to the 
presence of mass as the deviation from an otherwise flat plane according 
to (\ref{idil}). The TBH drags local spacetime with it in synchrony with 
the event horizon. Accordingly, inertial test particles travelling within 
the equatorial plane along initially radial geodesics from spatial 
infinity are compelled to orbit the TBH until their angular velocity 
reaches that of the event horizon. This occurs at the moment the horizon 
is crossed. Colour is used to denote the angular velocity of locally 
non-rotating observers as measured by distant observers, colours of longer 
wavelengths representing angular velocities approaching that of the TBH. A 
section of the outer funnel has not been plotted to provide visibility of 
the TBH aperture region. The ergoregions have not been identified here.

\section{JET FORMATION FROM TOROIDAL BLACK HOLES}

    The near-maximally rotating TBH undergoing accretion provides an 
excellent mechanism for the formation of ultra-relativistic (Lorentz 
factor $\sim10$) bi-directional jets as have been observed in quasars. The 
purpose of this section is not to explore the behaviour of the jets as 
they travel towards the distant radio lobes, the magnetohydrodynamics of 
which has been studied in great detail elsewhere, nor to analyse the 
myriad of particle interactions capable of extracting rotational energy 
from the TBH ergoregion. Rather, the essential differences between 
existing models and the accretion of matter onto a rotating TBH shall be 
outlined. Supermassive BHs have long been thought to reside at the heart 
of quasars and active galactic nuclei. Though masses as large as 
$10^9\msun$ have been associated with these, a typical rotating galaxy 
possesses ample angular momentum to spin up BHs of this size to maximal 
rotation. An upper limit on rotational velocity exists because the peak 
velocity of the event horizon cannot exceed that of light. In practice, 
the maximum rotation rate will not be reached, not least because the 
internal singularity would be revealed. Transportation of angular momentum 
away from the TBH by jets imposes more practical limitations. Hence the 
formation of jets, an intriguing feature of many AGN, is of paramount 
importance. Potential mechanisms responsible for their origination are now 
presented within the framework of the TBH model.

    A nascent galaxy may harbour a TBH whose spin rate is increasing. Once 
the spin reaches a plateau after a short delay (in cosmological terms), 
equilibrium is achieved and the accretion process is balanced by the 
angular momentum released by the TBH due to gravitational radiation, the 
production of jets and growth through capture of mass and angular 
momentum. Of these, the outflow of angular momentum is typically dominated 
by jet generation processes. This maintains a rapidly rotating TBH, but 
implies that accreting matter rotates with greater angular velocity than 
the spacetime near the TBH. Essentially, the interaction between this 
accreting mass and the enormous flywheel of the rotating TBH constitutes 
the basis for jet energy release. Fig.~6 illustrates a rotating TBH 
surrounded by an accretion disk. Since the TBH is able to shed any excess 
angular momentum by several mechanisms, its angular velocity is suppressed 
relative to the accretion disk $\omega_{tbh}<\omega_{disk}$. Apart from 
gradual mass accumulation, one can picture the TBH as being largely 
unaffected during periods of sustained activity, acting somewhat like a 
catalyser for the expulsion of angular momentum along the jet axis. Jet 
formation can therefore progress for substantial periods of time: 
$10^6\sim10^9$yrs.

\begin{figure*}
\centerline{\psfig{figure=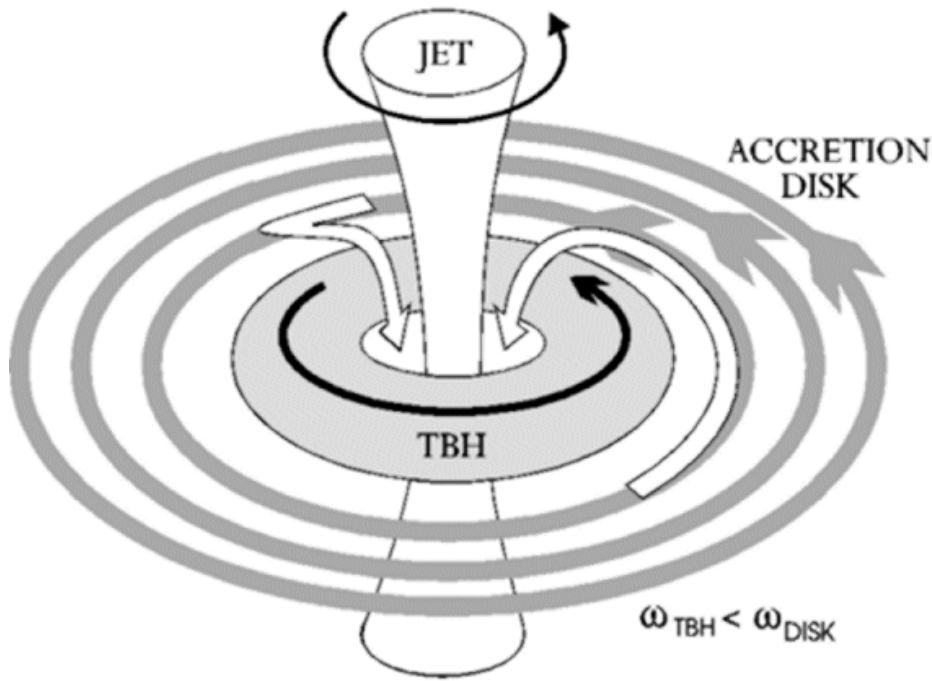,width=0.7\textwidth,angle=0}}
\caption{The transport of angular momentum in AGN. Rotational energy is 
supplied by the accretion disk to the TBH which is efficiently 
redistributed along axial jets by magneto-rotational mechanisms 
maintaining (i) the angular velocity differential between the TBH and 
accretion disk and (ii) the net charge on the TBH.}
\end{figure*}

    Within the central aperture, spacetime is dragged in concordance with 
TBH. The central aperture is a negative gravitational potential well, a 
spacetime vortex containing deeply negative energy states. Matter 
negotiating this central aperture will be obliged to travel along 
geodesics which appear to the external universe to be rotating. Particles 
capable of escaping to infinity require relativistic velocities closely 
aligned to the axis of rotation. Matter is able to travel in either 
direction along the rotational axis in order to achieve this, and angular 
momentum is transported away from the TBH equally by each jet. Of primary 
interest is a TBH whose central aperture is sufficiently small to provide 
powerful, collimated jets.

    Matter travelling through the aperture will undergo gravitational 
slingshot and could be propelled outwards along the jets, however the 
importance of this will be to impart kinetic energy leading to frictional 
heating. Temperatures of at least several million degrees (and perhaps as 
much as $10^{11}\degr$K) will be realised in the aperture, transforming 
the contained matter into a plasma emitting X-rays and some gamma rays.

\begin{figure*}
\centerline{\psfig{figure=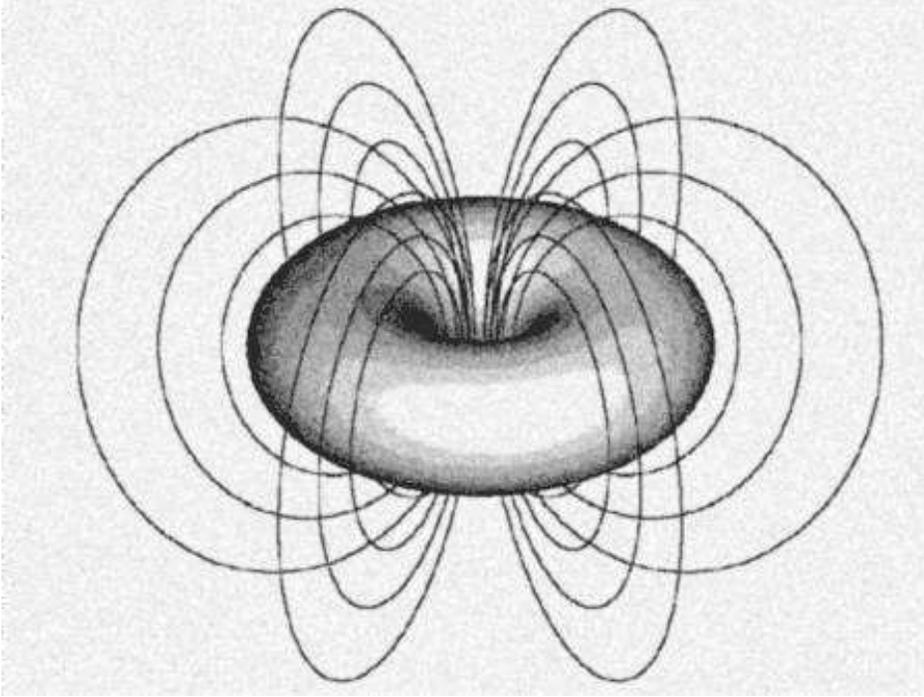,width=0.7\textwidth,angle=0}}
\caption{The magnetic field originated by a charged, rotating torus. The 
topology ensures that lines of magnetic flux never intersect the surface - 
therefore individually charged particles of the accretion disk spiral 
along the flux into the central aperture without neutralising the torus - 
in contrast, charged spheroids are easily neutralised since flux always 
terminates at the surface, typically in polar regions.}
\end{figure*}

    Ejection mechanisms such as the Penrose process (Penrose 1969), 
superradiant scattering (photonic counterpart of the Penrose process) and 
their analogues (e.g. due to particle-particle collisions) could dominate 
in the formation of jets. For convenience, the term Penrose process is 
used loosely to refer to all variations. The Penrose process exploits the 
existence of negative energy orbits inside the ergoregion of a rotating 
BH, permitting the extraction of energy to infinity at the expense of the 
rotational/kinetic energy of the BH. A particle travelling through the 
ergosphere might disintegrate into two particles, one of which plunges 
headlong towards the event horizon whilst the other emerges from the 
ergosphere and escapes to `infinity', e.g. as part of a jet. Energy is 
extracted if the emergent particle fragment has more energy than the 
originally intact particle, with the captured fragment carrying negative 
energy into the BH. The Penrose process efficiency improves if the 
particles have relativistic incident velocities, particularly those 
opposing the BH's rotation. A physical example of particle disintegration 
occurs within a high energy plasma when neutral hydrogen atoms are 
stripped of their electrons. More generally, ergoregional 
particle-particle collisions in which angular momentum and total energy 
are conserved may cause one of the resultant particles to be ejected to 
infinity (Piran, Shaham \& Katz 1975; Piran \& Shaham 1977). Particles are 
always ejected in a way that reduces the BH's angular momentum and 
rotational energy which increases the surface area of the event horizon, 
in keeping with the entropy law. As has been described by Wagh, Dhurandhar 
\& Dadhich (1985) and Bhat, Dhurandar \& Dadhich (1985) the presence of an 
electromagnetic field and/or charged particles can dramatically increase 
the efficiency of the Penrose process, easily to a level where net 
rotational energy may be extracted from the BH.

    The astrophysical significance of the Penrose process has been 
traditionally questioned, partly because BHs of spherical topology are not 
expected to retain significant electrical charge. It is argued that tori 
exhibit a vital difference. When a rotating torus accumulates charge, the 
circulating current establishes a poloidal magnetic field. Lines of 
magnetic flux encircle the torus but nowhere intersect its surface. Nearby 
the surface, flux lines are orthogonal to the current flow and parallel to 
the surface itself. Sufficiently intense magnetic fields constrain the 
motion of accreting plasma, obliging its constituent particles to follow 
helical trajectories which wind about lines of flux. Whereas charged 
spheroids are rapidly neutralised by plasma guided directly towards the 
surface by flux lines, neutralisation of charged tori is strongly 
inhibited due to the absence of flux lines intersecting the surface.

    According to the membrane paradigm, one can imagine the TBH's 
(infinitesimally stretched) event horizon to be an electrically conducting 
surface where electric fields incident to this membrane are terminated by 
an appropriate surface electric charge density. Also, the surface current 
will be such that the magnetic field parallel to the surface is 
terminated, in this way there will be no parallel magnetic field inside 
the event horizon. The effective surface resistivity will be of the order 
of several hundred ohms. The accreting matter contains neither magnetism 
nor net charge initially. The TBH is, however, spinning and dragging local 
spacetime around with it. The charged particles entering the ergoregions 
are mainly electrons and protons. Low efficiency Penrose processes will 
preferentially eject particles of larger charge/mass ratios (the 
electrons) and a net positive circulating charge will emerge hovering 
above the horizon. The toroidal membrane rotates and drags these positive 
charges around with it thus forming a circular electrical circuit. Current 
flowing in the circuit gives rise to an axial magnetic field through the 
central aperture, modulating the efficiency of particle emissions via the 
Penrose process thereby reinforcing the circulating charge and magnetic 
field. This magnetic field also plays a role in collimating the jets as 
they are launched, with charged particles spiralling along the magnetic 
field lines generating synchrotron radiation. Similarly, the dipolar 
magnetic field of the TBH channels free charged particles from the outer 
accretion disk into the central aperture, spiralling along the lines of 
magnetic flux.

\begin{figure*}
\centerline{\psfig{figure=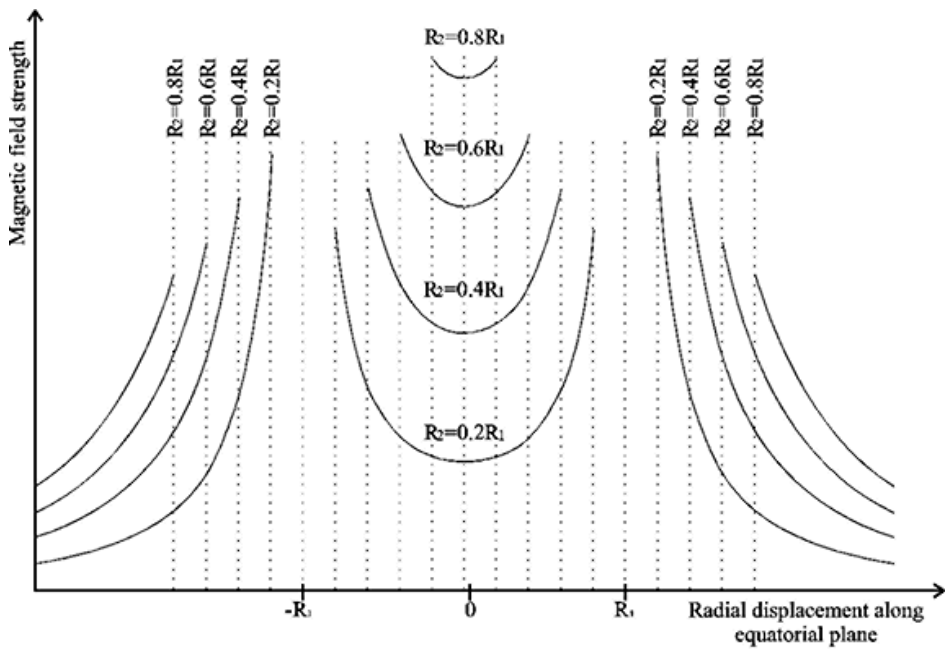,width=0.7\textwidth,angle=0}}
\caption{{\bf a)} Magnetic flux density in the equatorial plane of a 
rotating, charged toroidal shell for four separate geometries. The flux 
density can become arbitrarily large within the central aperture as the 
minor radius of the torus approaches the major radius.}
\end{figure*}

    The structure of the magnetic field encompassing a conducting toroidal 
shell is shown in Fig.~7. Lines of flux are illustrated which arise when a 
current circulates around the toroidal shell. Because the aperture can 
become arbitrarily small, if the total charge of the torus remains 
constant, the flux density and hence magnetic field can become arbitrarily 
large within this region. Computed plots of magnetic field strength along 
the equatorial plane of the torus are given for four separate toroidal 
geometries in Fig.~8(a). These have been calculated using the usual 
Biot-Savart relations and assume a constant and uniform surface current 
density $J$. Analytically, the magnetic field strength perpendicular to 
and within the equatorial plane at some displacement from the axis $a$, 
making use of symmetry, is given by the double integral:
\begin{gather}
\label{bofa}
B(a)\!=\!\!\int_0^{2\pi }\!\!\!\!\int_0^{2\pi}\!\!\!\!\!\!\!\frac{\mu 
_0Jt(a\cos\phi-t\cos{2\phi})}{
4\pi(a^2\!+\!t^2\!+\!R_2^2\sin^2\!\vartheta\!-\!2at\cos\phi)^{3/2}}
\,d\phi\,d\vartheta\\
\text{where\quad}t=R_1+R_2\cos\vartheta
\end{gather}

The Biot-Savart law simplifies at the centre of a circular current loop 
carrying a current $I$ and it is straightforward to verify that the field 
strength at that point is:
\begin{align}
B(0)=\frac{\mu_0I}{2R_1}
\end{align}

    Nearby the toroidal surface, Ampere's circuital law (\ref{acl}) states 
that the current enclosed by a closed path determines the sum of the 
magnetic field along the same closed path so the field strength is always 
finite at the shell's surface.
\begin{align}
\oint Bds=\mu_0\times I
\label{acl}
\end{align}

    The same law demonstrates that the integral (\ref{bofa}) is 
independent of $R_2$ providing $a<R_1-R_2$ or $a>R_1+R_2$, therefore some 
simplification is available by setting $R_2\to0$ whilst a constant current 
circulates. The integral of (\ref{idil}) can be expressed in terms of 
multiple elliptic integrals. Numerical computations have been used to 
derive plots in Fig.~8(a) which show that the field within the central 
aperture is generally stronger than in the outer periphery of the torus as 
measured by local inertial observers. This is particularly true for the 
geometries where $R_2\to R_1$ that would produce tightly collimated jets. 
The charged particles spiral around the strong field lines of the aperture 
achieving high velocities and alternate between contra-rotation and 
co-rotation during each cycle of their spiral. During the contra-rotation 
phase, they are especially likely to participate in ergoregional particle 
collisions in which energy and momentum is transferred to the jets at the 
expense of the angular momentum of the TBH.

    To better approximate the magnetic field of a charged TBH, the 
gravitational time dilation must be taken into account. Restricting 
analysis to the time dilation generated by a momentarily static ring 
singularity whose event horizon is transiently toroidal, the horizon must 
coincide in the equatorial plane with the surfaces of the electrically 
conducting toroidal shell. Essentially, this is achieved by precise 
adjustment of the radius of the ring singularity ($R_{ring}>R_1$) and the 
ring singularity's mass per unit length. The magnetic field strength plots 
of Fig.~8(a) are then recalculated, this time taking account of the local 
time dilation (lapse function) relative to a distant observer. The 
situation is analogous to earth based quasar observations because this 
magnetic field directly modulates the non-thermal radiation emanating 
mainly from within the TBH ergoregion. Note also, the negative energy 
states within the ergoregion of the central aperture will be more negative 
than elsewhere within the ergoregion but that the increased time dilation 
counteracts this. Results are presented in Fig.~8(b). Again, the 
electrical surface current density $J$ has been held constant but the TBH 
mass is different in each case.

    Several factors need to be considered which potentially impact the 
results. Time dilation  diminishes more rapidly than flux density with 
axial displacement. Improved models would involve non-uniform surface 
current densities for the TBH membranes, a means of estimating the TBH 
charge and a more accurate determination of the TBH angular velocity, 
which may require a theory of quantum gravity.

\begin{figure*}
\centerline{\psfig{figure=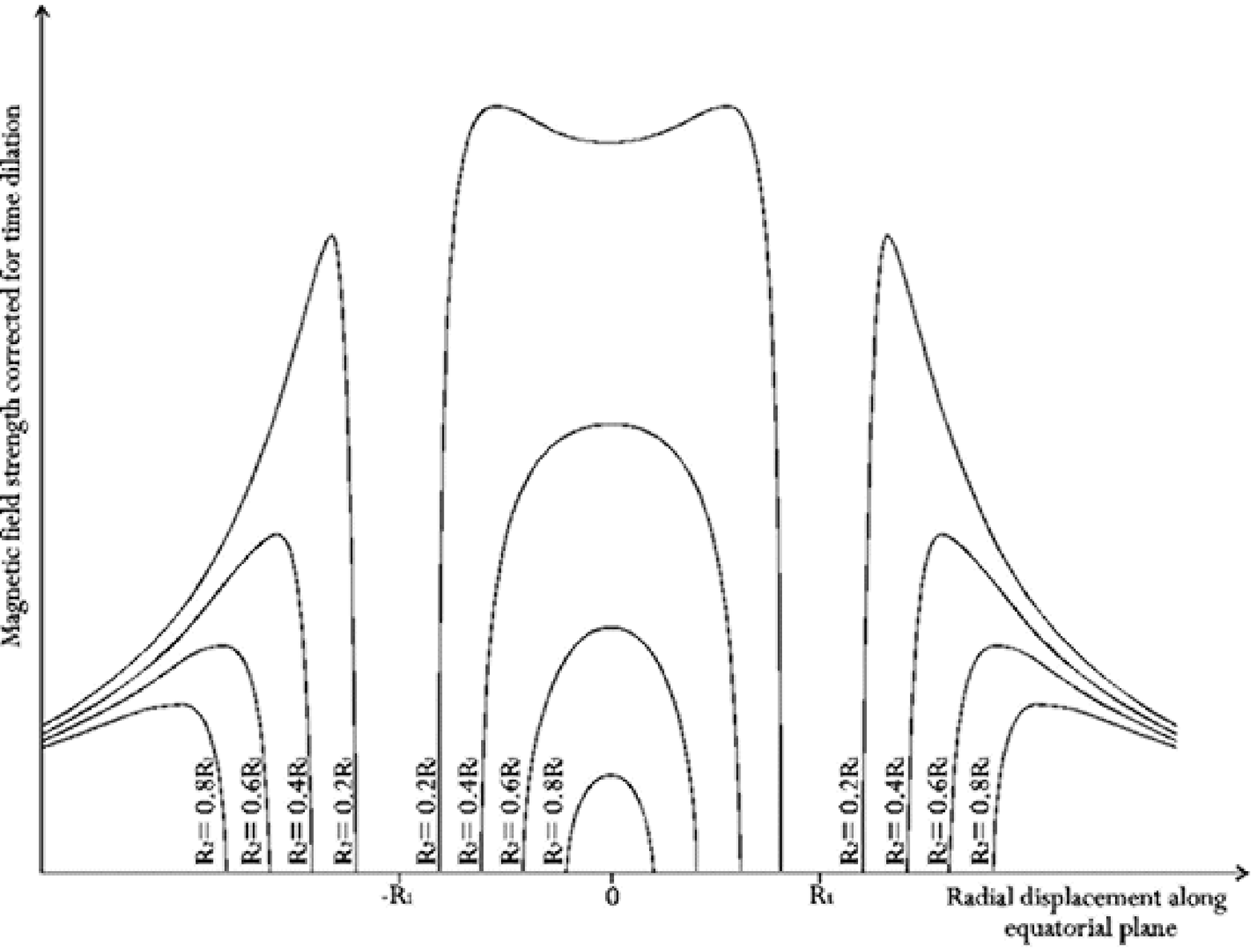,width=0.7\textwidth,angle=0}}
\setcounter{figure}{7}
\caption{{\bf b)} Magnetic flux density corrected for time dilation in the 
equatorial plane of a conservatively under-rotating, charged TBH for four 
geometries. More realistic, rapid rotation does not affect the time 
dilation along the axis of rotation but can boost the time dilation 
elsewhere, resulting in maximal effective flux densities located in the 
central aperture for all cases.}
\end{figure*}

    These facts imply that the Penrose process will occur predominantly 
within the central aperture of the TBH, and less so in the outer regions. 
Ergoregional particle emissions in the outer regions are largely 
reabsorbed by interactions with the accretion disk, whereas the rarefied 
central aperture allows relatively unimpeded passage to scattered 
particles. Thus, the most visibly energetic TBHs will be those with 
tightly focused jets.

    As discussed previously, because charge neutralisation is inhibited 
for tori, intrinsically stronger magnetic fields are to be anticipated in 
the vicinity of a TBH as compared to the rotating spheroidal BH situation. 
The presence of the magnetic field and the plasma gives rise to a 
force-free magnetosphere within the TBH's central aperture providing the 
plasma is sufficiently rarefied. The accumulation of circulating 
positively charged particles moderated by time dilation nearby the TBH 
event horizon is, for present purposes, identical to the situation where 
the TBH itself is charged. Supposing that the equilibrium magnetic field 
stabilises at large values $\sim10^{12}$\,G or larger, then vacuum 
breakdown may play a part in the formation of jets. Energy stored in the 
TBH magnetosphere would then be tunnelled quantum mechanically, creating 
pairs of charged particles and anti-particles. These virtual particles 
would then be separated by the intense electromagnetic forces before they 
could silently annihilate one another. The detection or otherwise of a 
significant positron population in the jets is a useful tool for resolving 
the issue of whether vacuum breakdown has a role to play.

    Energetic photons (X-rays and gamma rays) generated by the plasma of 
the central aperture would be emitted in all directions. Ergoregional 
processes would be capable of promoting them to higher energies because 
they are travelling at the speed of light, often in retrograde 
trajectories about the TBH. Resulting photons could inhabit the gamma ray 
spectrum at energies as high as the TeV range. The Penrose process should 
also act on the high velocity electrons and atomic nuclei of the 
high-energy plasma occupying the ergoregion of the aperture. At the 
expense of the rotational energy of the TBH, bi-directional jets with 
relativistic velocities are therefore likely to result. Matter and 
radiation must necessarily emerge from either side of the central 
aperture, where it begins its journey along one of the two jets. Depending 
on the geometry of the TBH, the jets could be tightly focused and 
penetrating or conversely, spluttering weakly over a broad solid angle. 
Particles are preferentially ejected in close alignment with the 
modulating magnetic field, in this case along the spin axis of the TBH. 
The toroidally originated magnetic field provides for deeper negative 
energy states within the ergoregion whilst extending the region of 
occurrence well beyond the static limit surface. Bhat, Dhurandar and 
Dadhich demonstrated that when charged particles are involved in Penrose 
process interactions there exists virtually no upper bound on the 
efficiency of energy extraction.

    The jets are to some extent able to collimate themselves if they are 
sufficiently focused at the source by trapping the magnetic field 
internally. Magnetohydrodynamic studies have had much success in 
explaining the characteristics of these outflows which emerge 
supersonically and travel for several millions of light years before 
ploughing into radio lobes formed at the ICM/jet interface. The knots 
frequently visible along jets are readily interpreted as the result of 
substantial short-term matter ingestion from stellar collisions with the 
TBH or instabilities in the accretion flow, one further possibility being 
that these knots may also be related to rapid fluctuations in TBH charge 
and instabilities in the TBH mechanism itself. The rotating jets will 
cause a net outflow of angular momentum from the TBH, which is 
counterbalanced by the net inflow of angular momentum due to accretion 
around the TBH's periphery. Jets transport angular momentum from the TBH 
because particles ejected from the ergoregions are rotating with the TBH 
having been launched by the Penrose process within the ergoregion, thereby 
generating a decelerational torque (recoil) on the TBH. By the mechanisms 
described, a significant portion of the mass and kinetic energy of 
accreting matter and radiation is available for jet production. For 
detailed analysis, numerical simulations will be required.

    The Penrose process reaches maximum efficiency when one of the 
particles heads directly towards the event horizon along the shortest path 
(i.e. it has the most negative energy state possible). Similarly, when the 
negative energy state arises due to the presence of charge on a particle, 
one of the particles emerging from the collision ideally scatters towards 
the event horizon along the shortest path. When the trajectory of the 
other scattered particle is considered for the purely gravitational 
Penrose process, the particle will head directly away from the event 
horizon, which for the TBH central aperture is typically a poor escape 
route. The situation is altered for electromagnetically dominated Penrose 
process interactions as the potential of the charged particle within an 
electric field should be considered. In order to recoil with maximal 
energy extraction, the charged particle will follow a path that leads 
towards greatest electrical potential which, for the aperture of a charged 
and rotating TBH, is aligned axially with the magnetic field. These 
ejected particles will frequently collide with the accretion flow 
streaming from the outermost periphery of the TBH. The jets will be 
sufficiently strong to overcome this inward accretion flow in the regions 
nearest the spin axis. An almost identical scenario was analysed 
(Blandford \& Rees, 1974) wherein hot, relativistic plasma escapes 
anisotropically through orifices punctured in a cool surrounding gas 
resulting in beams of collimated plasma.

    Data gleaned from quasar observations is consistent with the present 
TBH model. High energy gamma rays at energies up to $20\,$GeV $\sim 
1\,$Tev have been detected within the jets. These may correspond to 
photons ejected by the Penrose process, rather than by some secondary 
acceleration mechanism within the jets. The variation in jet dispersion 
angles is related to the $R_2/R_1$ ratio of the TBH and is naturally 
accounted for by the opening angle of the TBH. Quasar spectra can contain 
three separate red-shifted portions; the TBH model displays a similar 
complexity:

\begin{itemize}
\item
The plasma of the central aperture is buried in a deep gravitational 
potential well.
\item
Jets travel relativistically in opposite directions, one of which is 
usually not directly detectable.
\item
Radiation passes through the metal enriched clouds generated by the SN of 
the TBH creation event.
\item
The remoteness of the QSO galaxy correlates to a cosmological recession 
and red-shift.
\end{itemize}

    A carefully considered numerical study of gravitating fluids (Marcus, 
Press \& Teukolsky 1977) reveals a bifurcation from the Maclaurin 
ellipsoids to lower energy state 'Maclaurin toroids' at high angular 
momenta which the authors suggest may be stable against all small 
perturbations. Alternatively, toroidal density distributions (TBHs or 
transient neutron tori) may develop in the dynamically collapsing cores of 
moderately rotating progenitor stars (rotary core collapse). Butterworth 
\& Ipser (1976) demonstrated that ergoregions can form when relativistic 
stars spin rapidly although absolute event horizons are absent. With the 
notable exception of long-term stability, charged neutron tori could share 
many similarities with the charged TBH central engine of quasars. Although 
the neutron torus could exhibit various non-axisymmetric instabilities, it 
appears that the electromagnetic structure could significantly counteract 
these effects for core-collapse timescales if electrical charge gathers on 
the torus. Moreover, stability will clearly be reinforced if the 
composition of the torus becomes superfluidic and superconducting, as is 
likely for the least soft equations of state.

\begin{figure*}
\centerline{\psfig{figure=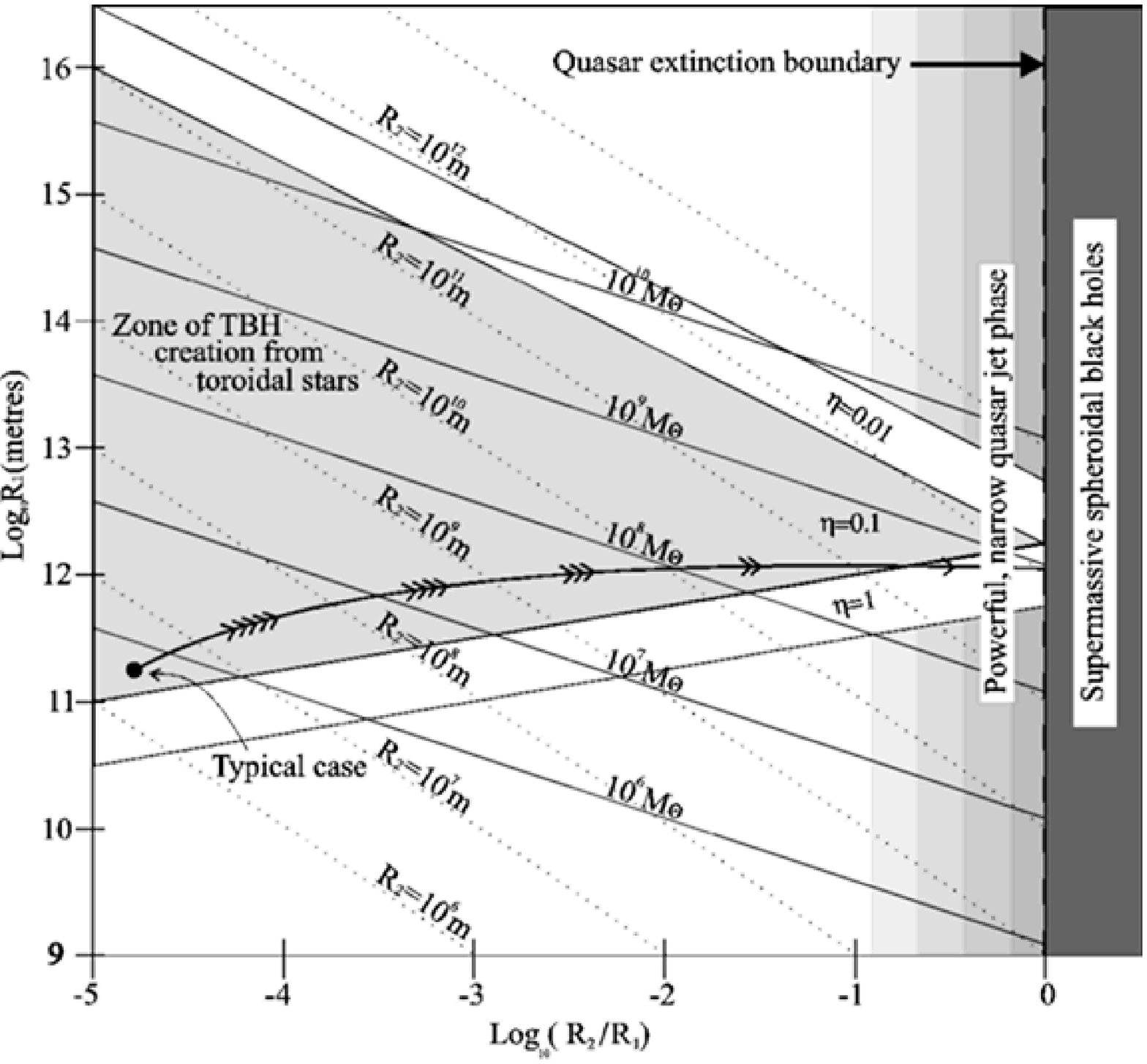,width=0.7\textwidth,angle=0}}
\caption{Toroidal black hole creation, evolution and extinction. Contours 
of constant mass can only be traversed in one direction. A typical case is 
identified which corresponds to the lifecycle of a prototypical quasar. 
AGN extinction accompanies the topological transition at $\log(\frac{R_2}{
R_1})=0$. Beneath the shaded wedge, TBH formation can occur from the 
collapse of degenerate tori.}
\end{figure*}

    Rotating, electrically charged neutron tori are able to generate 
immensely strong axial magnetic fields. When ergoregions arise and 
accreting matter is available, relativistic bi-directional jets should 
emerge. Given a steady supply of mass and angular momentum, as from an 
accretion disk fuelled by a binary companion star, microquasar behaviour 
could result --- though long-term stability arguments favour a TBH engine. 
Certain explosive core-collapse SNe (hypernovae) accompanied by 
anisotropic gamma-ray bursts (GRBs) could be interpreted as the outcome of 
a neutron/BH torus forming during rotary core collapse. Metastable 
configurations can be envisaged in which a white dwarf accretes matter 
from a companion star, the angular momentum accumulates until the core 
eventually collapses to a neutron torus causing a brief but intense 
outburst of mass and angular momentum sufficient for the star to resume 
its quiescent spheroidal white dwarf state. Recovery from neutron to white 
dwarf densities is permitted, but event horizons grow irreversibly. Rotary 
core collapse may curtail the birth of TBHs at masses substantially 
$below$ the lower limits commonly adopted for BHs.

\section{EVOLUTION BEYOND OPTICALLY BRIGHT QSO PHASE}

    It is well documented that quasar populations decline rapidly at 
co-moving red-shifts below $z\sim2.5$. What is not easily explained by 
leading models reliant on a central supermassive spheroidal BH is why 
quasar activity terminates so abruptly in recent times. This is especially 
puzzling given that the BHs themselves cannot have altered other than gain 
yet more mass from their surroundings. Ideas have been proposed such as 
advection dominated accretion flow (ADAF), the `spin' paradigm and various 
`state transitions' related to accretion efficiency. A significant 
diminution of BH angular momentum is unlikely given the tendency of 
accretion disks to transfer rotational energy to a BH. If fundamentally 
different modes of accretion do operate, lower efficiency modes will be 
obliterated whenever a massive body approaches the disk/engine 
environment. Mass injection of this kind would sporadically re-establish 
brief quasar behaviour in previously dormant galaxies, including 
neighbouring galaxies which are observed to be emphatically inactive. 
Evidence for such transient activity is absent both in local and distant 
galaxies. This is exceedingly troublesome for standard AGN models when 
accounting for the quiescent cores of nearby giant ellipticals.

    According to the present model, the TBH will transition to a 
spheroidal BH once accretion has inflated the event horizon or decreased 
the angular momentum sufficiently. This provides a natural mechanism for 
the termination of quasar-like TBH activity within the universe and is 
amply supported by observations. The swelling of the toroidal event 
horizon due to mass capture generally overcomes the increasing angular 
momentum of the TBH by the same process. Although the major radius of the 
torus may be increasing, the minor radius will eventually catch up leading 
to a topological transition. Immediately prior to the extinction of the 
TBH, the most energetic and tightly collimated jets are anticipated to 
form, albeit with enhanced gravitational red-shifts as seen from infinity.

    Fig.~9 illustrates a number of features of the TBH quasar model and 
has been constructed using the inequality relations (12) to (15) from 
section 2. The shaded wedge represents the area within which TBHs can come 
into being directly from the implosion of a toroidal star. Here, it has 
been assumed that the seed star has constant density of 
$\sim1400\,$kg\,m$^{-3}$ and that 90\% of the star's mass is ejected 
during the implosion ($\eta=0.1$). TBH creation at radii $R_1$ below about 
$36\times 10^9\,$m is prohibited because electron or neutron degeneracy 
would halt the collapse, as it would slightly beyond the left hand edge of 
the diagram at about $R_2/R_1\sim10^{-6}$ and beyond, so that the wedge 
shape does not continue indefinitely. Above the shaded wedge, it is 
impossible for the toroidal star to be sufficiently massive if, as it must 
be, $R_3<R_1$ for the seed star and densities above $1400\,$kg\,m$^{-3}$ 
are disallowed. Lightly shaded regions above and below the main wedge show 
how the diagram would be altered if different values were taken for 
$\eta$.

    This diagram identifies the region at which quasar-like behaviour is 
to be expected from TBHs (progressively shaded vertical section) where 
$R_2$ is almost as large as $R_1$ and narrow jets are formed. The line 
defined by $R_2=R_1$ is the quasar extinction boundary where the TBH 
becomes a spheroidal BH. Lines have been plotted to indicate contours of 
equal TBH mass and similarly for constant minor radius $R_2$. Because the 
mass of the TBH will not diminish with time, constant mass boundaries can 
only be traversed in one direction. There is a sufficiently broad birth 
zone spanning several orders of magnitude on each axis which enhances the 
probability that TBH creation is widespread at the centre of typical 
protogalaxies.

    TBH birth is anticipated to occur at lower masses and lower $R_2/R_1$ 
ratios because the seed stars required are very large even for these. One 
typical case has been presented on the diagram. For this example a 
toroidal star of mass $6\times10^6\msun$ and radii $R_1=1.8\times10^{11}$m 
and $R_3=8.7\times10^{10}$m implodes after exhausting its fuel on a very 
short timescale to leave a TBH of mass $6\times10^5\msun$ and radii 
$R_1=1.8\times10^{11}$m and $R_2=2.8\times10^6$m. The accretion rate 
within this young galaxy is increasing so the TBH mass swiftly increases 
as does its angular momentum and angular velocity. The arrows of the 
evolutionary trace depict the evolutionary rate, fastest at the start then 
slowing down at higher masses such that the QSO phase can exist for a 
timescale several orders of magnitude greater than the formation time of 
the TBH. Somewhat inevitably, when $R_1$ and $R_2$ equalize, the quasar 
phase is discontinued, in this example when the mass reaches about 
$10^9\msun$. For a given mass, the relationship between $R_1$ and $R_2$ 
will depend upon the angular velocity of the TBH which in turn is related 
to the angular momentum inflow due to accretion. In order to sustain a 
toroidal event horizon indefinitely, an ever increasing supply of angular 
momentum may be required. The accretion rate might be relatively low at 
the time of TBH birth, rising swiftly before peaking and slowly decreasing 
thereafter. Accordingly, the jet formation phase is predicted to terminate 
due to the topological transition at the boundary where $R_2\to R_1$.

\section{DISCUSSION}

    It has been qualitatively described how rotating TBHs might evolve 
from protogalactic gas clouds and accrete matter from the galactic centre 
until their inner apertures contract and highly focused relativistic jets 
form. The viability of the model can be tested by observing the evolution 
of jet collimation with red-shift as this model predicts the degree of 
collimation continues to increase (though jet energetics may decline at 
later times) until the topological transition. Such an approach could 
further address the issue of whether the entire AGN population or some 
subset thereof is accounted for by a TBH model. The direct observations of 
AGN and quasars in our universe suggests that TBHs are more than abstract 
mathematical constructs. Classical general relativity still remains to be 
unified satisfactorily with quantum mechanics. Evidently, TBH stability is 
intertwined with this issue. As direct experiments cannot be performed in 
intensely curved spacetimes, astrophysical observations must be our guide. 
The proposed stability of TBHs allows the event horizon's interior to be 
metaphorically unveiled, providing clues to the nature of quantum gravity 
and grand unification theories.

    General relativity demands that the cosmological constant $\Lambda$ be 
sufficiently negative in order to provide long-term TBH stability. 
Observational estimates of $\Lambda$ based upon universal expansion 
reliant on the behaviour of general relativity in weak field environments 
suggest that its value is very small but probably positive. One 
possibility is that $\Lambda$ is primarily a function of local spacetime 
curvature. Alternatively, the presence of external matter (accretion disk 
and galaxy) may provide the necessary curvature and non-stationarity 
permitting TBH stability over indefinite periods.

    The TBH model, despite its controversial nature, presents a promising 
means of understanding the following characteristics of quasars: extreme 
jet energies, varying jet emergence angles, abrupt extinction, high 
gamma-ray radiation, the presence of heavy elements and the multiplicity 
of red-shifts in absorption spectra. None of these features are readily 
explained by spheroidal BH models. It is encouraging that the TBH model 
also appears to lead to plausible models for macroscopic processes within 
supernovae, microquasars and gamma-ray bursts. Fortunately, neutron tori 
-- unlike TBHs -- are exempt from topological censorship. Hence, these 
will certainly exist in astrophyscial circumstances, if only very briefly. 
Betraying their existence in distinctive and overtly energetic ways, these 
curiosities should be amenable to observational identification and study.

    Numerical simulations are crucial if accurate comparisons with further 
detailed observations are to be made. Gravitational wave detectors and 
planned optical/X-ray interferometer technology will be sufficiently 
advanced in forthcoming decades to conclusively resolve the question of 
whether toroidal black holes truly exist.

\section*{ACKNOWLEDGEMENTS}
I wish to thank my father, Andrew Spivey for numerous enlightening 
discussions and continued assistance. Thanks also to Antony Hewish of 
Cambridge University for initial comments.

\vfill \eject

\appendix{\centerline {\bf APPENDIX A} }

\renewcommand{\theequation}{A.\arabic{equation}}
\renewcommand{\thefigure}{A.\arabic{figure}}

\medskip
\noindent
    General relativity does not restrict the distribution of mass-energy 
external to event horizons. The question is not whether, but to what 
extent can toroidal arrangements of neutron degenerate matter be 
temporarily stabilised by rotation. The purpose of this appendix is to 
outline and prepare quantitative estimates of the circumstances leading to 
the creation of neutron tori using extensive simplifying assumptions due 
to the complexity of the situation. Although it seems that rather extreme 
conditions are necessary for tori to form, and dynamic evolution could be 
both violent and rapid, it will be seen that rotary core collapse 
supernovae are a natural setting for the birth of dense tori. Although the 
analysis presented is relevant to a broad range of stellar densities, 
including the formation of toroidal black hole cores, it must be stressed 
that tori of neutron density and below cannot be instantly dismissed by 
topological censorship - the primary objection to the formation of TBHs. 
It is known that ergoregions can form when relativistic stars form with 
high angular momentum despite the absence of either apparent or absolute 
event horizons. Potentially, this permits the proposed quasar mechanism to 
operate in a number of seemingly unrelated astrophysical phenomena.

    Many stars, particularly the brighter Type Oe and Be, rotate 
considerably faster than the Sun. Equatorial velocities in the range 
$300\sim700\,$km\,s$^{-1}$ are not uncommon as compared to $2\,$km\,s$^{-1}
$ for the Sun. It is thought that the majority of stars have high initial 
angular velocities but that coupling between solar winds, magnetic fields 
and the inter-stellar medium cause a gradual decline in angular momentum. 
The brighter, more massive stars can be very short-lived and will retain a 
large angular momentum once their nuclear fuel is spent. It is therefore 
worthwhile studying the internal structure of rapidly rotating stars 
undergoing gravitational collapse to determine the conditions best suited 
for producing toroidal core configurations.

    In order to preserve analyticity, a simplified model is presented. 
Later it will be apparent that removal of any simplifications necessitates 
a numerical treatment such as the one presented by Marcus, Press \& 
Teukolsky. A uniformly rotating (constant angular velocity) ellipsoid 
would consist of ellipsoidal shells for which the assumption of uniform 
shell density is invalidated in situations of interest. Instead, uniformly 
rotating infinite cylinders are first considered. A cross section is 
illustrated in Fig.~A.1(a) of the proposed structure of a rotating star 
composed of two immiscible, incompressible fluids with densities $\rho_1$ 
and $\rho_2$ where $\rho_2>\rho_1$. The contours denote lines of equal 
hydrostatic pressure increasing from zero at the surface of the spheroidal 
envelope to a peak within the higher density toroidal core.

\begin{figure*}
\centerline{\psfig{figure=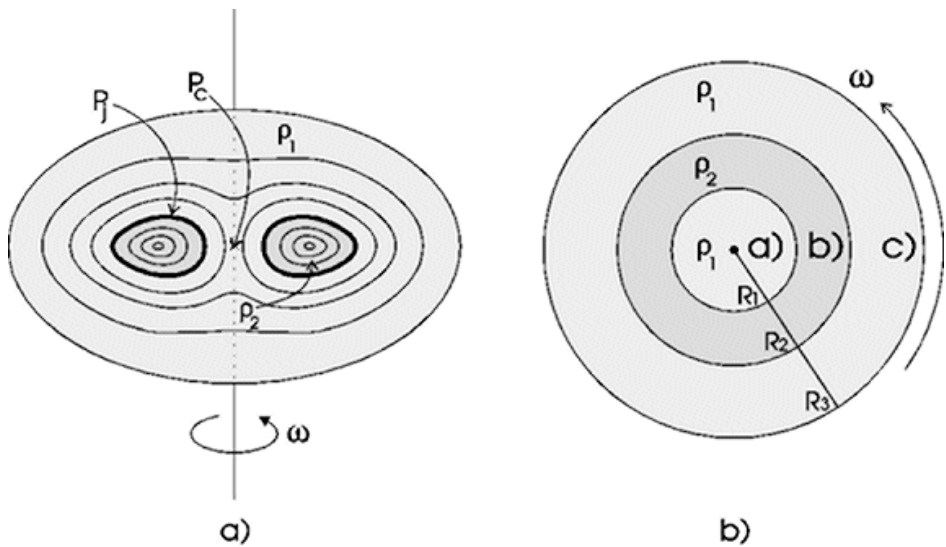,width=0.7\textwidth,angle=0}}
\caption{a) A rotating ellipsoid enveloping a denser toroidal core is here 
approximated by...
b) A rotating infinite cylinder (cross section illustrated) composed of 
two incompressible
fluids arranged in a pseudo-toroidal formation where $\rho_2>\rho_1$.}
\end{figure*}

    If the pressure and density within a cylinder composed of two fluids 
can be shown to reach a maximum at a finite distance from the axis of 
rotation, Fig.~A.1(b), and that such a distribution is in equilibrium, 
then the some degree of stability can be ascribed to the arrangement in 
Fig.~A.1(a). A Newtonian analysis permits exact solutions by virtue of the 
superposition of cylindrical shells of differing densities, the linearity 
of (constant density) cylindrical gravity with radius, the null gravity 
within infinite cylindrical shells and the gravitational equivalence of 
cylindrical shells to axial line masses in the exterior regions. 
Cylindrical coordinates $(r,\phi,z)$ are employed to consider two 
immiscible and incompressible fluids of densities $\rho_1$ and $\rho_2$ 
with $\rho_2>\rho_1$ rotating smoothly and uniformly. Temperature is 
neglected because degenerate materials are of particular concern. The 
gravitational profile is not linear with radius because it is caused by 
three zones of different density: regions a), b) and c) with radii $R_1$, 
$R_2$ and $R_3$ respectively indicated in Fig.~A.1(b). When equilibrium is 
achieved, the resultant force on each fluid element is zero. In this 
analysis, the individual forces acting on the elements are due to the 
pressure gradient, the centripetal acceleration and the gravitational 
attraction which, by virtue of the cylindrical symmetry, need only be 
considered in the radial direction. It is elementary to derive expressions 
for the derivatives of pressure, $P_a, P_b$ and $P_c$, with respect to 
radius for each of the three regions a), b) and c) respectively:
\begin{align}
\label{press}
\frac{dP_a}{dr}&=r\rho_1(\omega^2-2\pi G\rho_1)\\\frac{dP_b}{dr}
&=r\rho_2\left\{\omega^2-2\pi G\left[\rho_2+\left(\frac{R_1^2}{r^2}
\right)(\rho_1-\rho_2)\right]\right\}\\
\frac{dP_c}{dr}&=r\rho_1\left\{\omega^2\!-\!2\pi 
G\left[\rho_2\!+\!\left(\frac{R_1^2-R_2^2}{ r^2}\right)(\rho_1\!-\!\rho 
_2)\right]\right\}
\end{align}

    These expressions are readily integrated using the following boundary 
conditions: $P_c(R_3)=0, P_b(R_2)=P_c(R_2)$ and $P_a(R_1)=P_b(R_1)$. It is 
immediately apparent from (\ref{press}) that the pressure increases with 
radius from $r=0$ providing a certain minimum angular velocity is 
exceeded: $\omega>\omega _{\rm min}=\sqrt{2\pi G\rho_1}$. For physically 
meaningful results, the pressure must not become negative at radii 
occupied by matter. There are two circumstances where this might first 
arise: at the centre (when the cylinder becomes hollow) and immediately 
beneath the surface (centripetal forces overcome gravitational forces 
leading to mass shedding). The latter condition is simply expressed as 
$dP_c/dr>0$ at $r\to R_3$ permitting the definition of a maximum angular 
velocity $\omega_{\rm max}>\omega_{\rm min}$ which is conveniently 
expressed as:
\begin{align}
\omega_{\rm max}=\omega_{\rm min}\times\sqrt{1+\left(\frac{\rho_2-\rho_1}{
\rho_1}\right)
\left(\frac{R_2^2-R_1^2}{R_3^2}\right)}
\end{align}

    In general, a wide range of angular velocities are available if the 
densities are very dissimilar whereas, as the densities of the two fluids 
approach one another, there is a much narrower range of values that 
$\omega$ can occupy above $\omega_{\rm min}$. A specific example is given 
in which the radii are in the ratio 1:2:3 for $R_1:R_2:R_3$ and the 
densities 1:2 for $\rho_1$ and $\rho_2$. Results are plotted in Fig.~A.2.

    The diagram presents the pressure variation along the radius of a 
rotating infinite cylinder. Several curves have been plotted which 
correspond to different rates of rotation. For the example given, internal 
pressure remains positive up to $\omega\to1.155\omega_{\rm min}$, the mass 
shedding limit. The stability of these results is trivial because the 
assumption of equilibrium was inherent in the model, all solutions are in 
neutral equilibrium including those at low angular velocity and the 
non-rotating case. The significance of $\omega_{\rm min}$ is that 
stability cannot be achieved below this if the fluids become 
infinitesimally compressible because the density and pressure 
distributions would be qualitatively different.

    When a homogeneous rotating cylinder of compressible fluid is 
considered, it transpires that stability of off-axis peak density 
arrangements is unattainable if the assumption of uniform rotation is 
retained. This is evident upon inspection of (\ref{press}) for cases where 
the pressure and density are positively correlated i.e. $dP/d\rho>0$. This 
does not mean that an axial pressure-density peak will always result, once 
the central angular velocity exceeds a certain value $\omega_c\sim\sqrt{
2\pi G\rho_c}$ then axial density peaks are also unstable and differential 
rotation will occur (non-constant angular velocity). Furthermore, the 
angular velocity will generally decrease with radius for these systems, 
coinciding with the most physically realistic situations exemplified by 
the frame-dragging of the Kerr spacetime. Hence a non-axial density peak 
akin to toroidal solutions of rotating gravitating spheroids results. A 
numerical treatment must be employed for systems of even this complexity, 
before realistic interactions such as viscosity, radiation pressure, 
temperature variations and magnetic braking are incorporated in the model. 
Analytical limits of relevance to infinite cylinders exist in (i) the 
differentially rotating incompressible fluid approximation and (ii) the 
uniformly rotating compressible fluid regime whereby iterative solution of 
Volterra integral equations is in principle achievable. Numerical 
techniques become mandatory for differentially rotating compressible 
fluids in the infinite cylinder approximation.

    It should be stressed that long term stability is not the issue here 
-- all that is required is for a differentially rotating toroidal 
structure to transiently exist during the inherently dynamic and unstable 
collapse phase of stellar evolution - the necessary timescale for 
`stability' is briefer than that required for dissipative processes to 
restore uniform rotation, e.g. magnetic braking and viscosity.

\begin{figure*}
\centerline{\psfig{figure=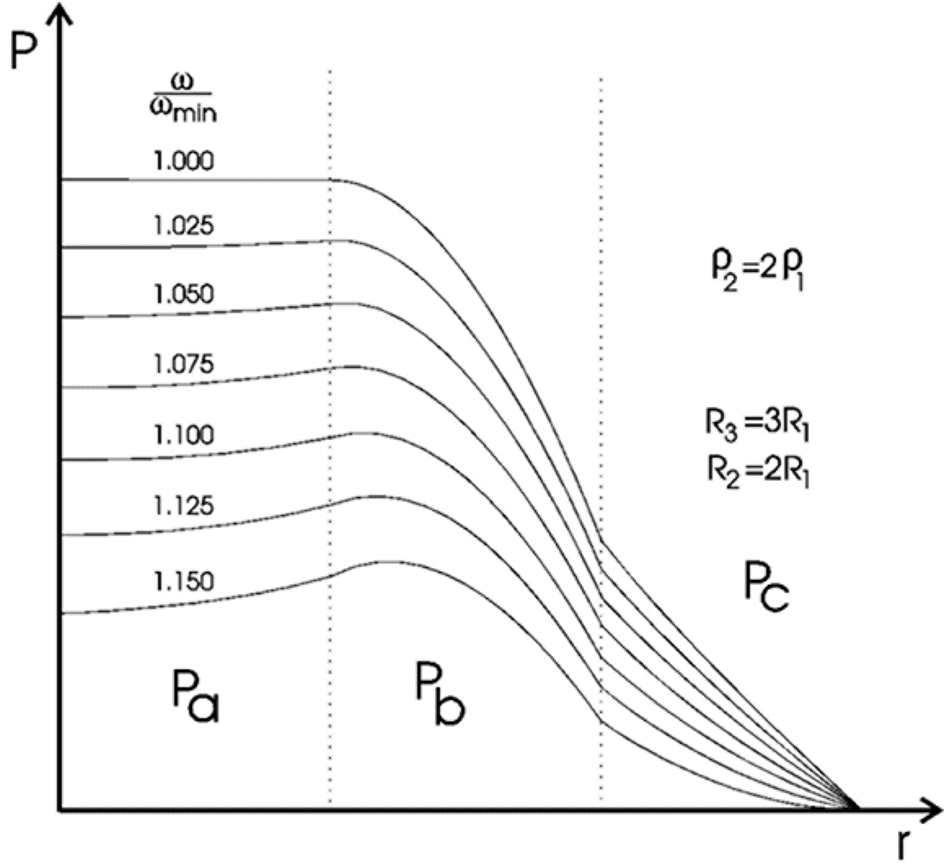,width=0.7\textwidth,angle=0}}
\caption{Pressure variation with radius for rotation rates of interest. 
Quasi-stability occurs for compressible fluids only when the pressure is 
maximal within region b). At higher rotation rates, mass is shed from the 
surface into keplerian orbits.}
\end{figure*}

    When a rapidly rotating star collapses, and the collapse originates at 
the core where gravity first defeats pressure, strong differential 
rotation accompanies a radial inrush of material - this environment 
facilitates the formation of a toroidal core. Because angular momentum is 
conserved for all collapsing shells, the angular velocity in the core 
increases appreciably to a level in excess of $\omega_{\rm min}$ and the 
angular velocity beyond the core declines radially. If the core is 
sufficiently dense, relativistic frame dragging contributes to the 
differential rotation. Radially decreasing rotation coupled to the fact 
that the pressure (and therefore the density) must increase with radius at 
the centre means that toroidal cores are practically inevitable during the 
collapse of rapidly rotating stars. Axisymmetric tori often exhibit 
non-axisymmetric instabilities in numerical simulations, the resultant 
gravitational radiation being of fascination to gravitational wave 
astronomy. The perturbational influence of orbiting companions could also 
disturb the symmetry. A simple equation of state is insufficient to model 
superfluidity and superconductivity, properties that neutron stars are 
widely expected to possess, rendering inapplicable many models constructed 
to investigate gravitational wave driven instabilities. The self-gravity 
of a torus can sustain a state of pseudo-equilibrium. If the torus is 
electrically charged, the exterior magnetic field acts as a barrier both 
to surface winds consisting of charged particles and to neutralising 
inflows -- reinforcing stability. The outcome for situations where both 
$\omega_{\rm min}$ and $\omega_{\rm max}$ is exceeded is qualitatively 
unchanged. Mass expelled from the outermost periphery of the torus is not 
ejected to infinity but forms an equatorial disk orbiting the torus. If 
sufficient mass is shed, this disk can become geometrically thick, i.e. 
the torus expands.

    For slowly rotating ellipsoids, the value of $\omega_{\rm min}$ 
increases to $\sim\sqrt{4\pi G\rho_c}$. If the neutron core following 
gravitational collapse of the Sun has a density in excess of 
$\sim~3\times10^{18}\,$kg\,m$^{-3}$ then it is conceivable that the core 
could become toroidal. If a more attainable collapse density of $10^{14}
\,$kg\,m$^{-3}$ were specified, then the Sun would only need to rotate at 
a moderately higher angular velocity of $\sim5.5\,\omega_{\rm sun}$ for 
$\omega_{\rm min}$ to be attained during core collapse. The Sun's internal 
pressure would first vanish within the surface (mass shed into keplerian 
orbit) were it to rotate at a rate $\sim212\,\omega_{\rm sun}$, so the 
available rotation range for toroidal core collapse is generally broad and 
attainable for typical massive stars.

    A TBH embedded in a collapsing stellar envelope resembles a 
scaled-down version of the described quasar environment. In contrast with 
the sustained activity of active galaxies, the tremendously accelerated 
accretion onto the core causes a violent and intense anisotropic 
explosion. Such a TBH could be of relatively low mass, perhaps below 
$0.1\msun$ as efficient jet generation processes begin to operate and 
stall its growth. There is therefore a definite possibility that a large 
fraction of the remaining star is propelled along the jets. Together with 
the first active galaxies, there are obvious implications for cosmological 
reionization which appears to have occurred at red-shifts of $z\sim6$. A 
neutron torus formed during rotary core collapse can `recover' following 
the ejection of substantial mass from the stellar envelope to a lower 
density object e.g. a white dwarf. On the other hand, charged neutron tori 
that do not recover will subsequently collapse to charged spheroidal 
neutron stars forming highly magnetised pulsars or Kerr-Newman BHs. 
Objects classified as `magnetars' or anomalous X-ray pulsars (AXPs) have 
been observed. The inferred magnetic field strengths due to the spin-down 
rate of these objects is $\sim10^{14}$G, providing a useful clue as to the 
degree of electrical charge of the neutron tori during SNe and thus, by 
inference, the net TBH charge in AGN circumstances. This level of charge 
can have a significant bearing on the spacetime geometry -- this is 
discussed in appendix~B.

    White dwarves containing toroidal neutron cores and isolated rotating 
neutron tori will often form during the core collapse of moderately 
rotating progenitors. The rotation rate should be sufficient for these 
dense tori to generate ergoregions thence accumulate charge sustaining 
magnetospheres which bolster the negative energy states of the ergoregion. 
These neutron tori are unlikely to be long-lived as they are susceptible 
to various instabilities and their differential rotation will eventually 
be erased by dissipative processes, but they are certainly of interest in 
more dynamic environments. The significance of these short-lived neutron 
tori or equivalently stellar mass toroidal black holes (SMTBH) are now 
addressed in three separate astrophysical settings.

    Firstly, a binary system consisting of a SMTBH and a stellar companion 
could operate as follows: an accretion disk forms around the SMTBH 
composed of material transported from the nearby star by gravitational and 
electromagnetic interactions. The central aperture of the SMTBH contains 
an ergoregion and an intense dipolar magnetic field due to a net 
electrical charge. This then gives rise to anti-parallel jets aligned with 
the axis of rotation in a very similar manner to the quasar albeit on a 
smaller scale. Microquasars have been observed within the confines of the 
Milky Way and are so called because they seem to obey simple scaling laws 
applied to quasars. Accretion of material from the companion star and jet 
formation will combine to decrease the overall angular momentum of the 
SMTBH on a shorter timescale than that of the quasars, the unusual 
behaviour terminating when the topology transitions to spheroidal after 
the angular momentum has been partially jettisoned. Unlike pulsars, the 
magnetic fields generated by charged tori will be robustly aligned with 
the rotation axis and, with the possible exception of radiation emanating 
from an accretion disk, periodic bursts of radiation will not be observed.

    Relativistic galactic jet sources and their similarities with quasar 
outflows have been reviewed by Mirabel \& Rodriguez (1999). From the 
limited microquasar observations available, it appears that the jet 
velocities have a bimodal distribution classified by $\nu_{jet}\approx 
0.3c$ and $\nu_{jet}\ga0.9c$. Whether a corresponding distribution exists 
for the jets of active galaxies is currently unknown. An interesting 
feature of some microquasars which is absent in quasars is their behaviour 
as the accretion disk is exhausted resulting in a sudden ejection of 
condensations (Mirabel et al 1998). Existing steady state MHD models with 
continuous jets have difficulty accounting for this, relying on a 
disk-supported magnetic field. This problem is resolved in the present 
model because the magnetic field of the torus remains when the disk 
disintegrates and confines the remaining plasma to circulate above the 
event horizon of the SMTBH, weaving repeatedly through the central 
aperture until it emerges in the form of jets aligned with the spin axis.

    Secondly, the mechanism could participate in the most energetic SNe - 
those which have been dubbed `hypernovae' with energies two orders of 
magnitude above `ordinary' SNe and thought to coincide with longer 
duration gamma-ray bursts. GRB980425 has been associated with SN1998bw 
providing evidence for a common mechanism (Cen, 1999). For example, 
consider a massive, rotating and fuel starved star undergoing rotary core 
collapse. A neutron torus (or SMTBH) develops in the core embedded within 
a lower density envelope. As before, charge accumulates on the torus which 
cannot be quickly neutralised on a timescale comparable to that of the 
implosion. A strong magnetic field threads the central aperture of the 
torus and strongly negative energy states are available in the ergoregion. 
The mechanism results in a ferocious outward explosion of matter from the 
centre of the SN in which a significant proportion of the star's mass is 
expelled anisotropically. Jets from SNe have been inferred from nearby 
hot-spots detected by optical speckle interferometry (Cen 1999, and 
references therein). Evidence of highly anisotropic ejecta is provided by 
polarimetric SN observations (Wang et al, 1999). Collapsar models 
attempting to account for jet formation in core collapse SNe are hampered 
by the spherical core topology as jets, if they form at all, are unlikely 
to penetrate through imploding shells for the following reasons:

\begin{itemize}
\item
The inefficiency of the Blandford-Znajek mechanism is now recognised.
\item
The topology dictates that plasma flowing along magnetic flux lines 
efficiently neutralises the core --- but in the case of toroidal cores, 
neutralisation paths are orthogonal to flux lines, and neutralisation is 
thereby inhibited.
\item
The equatorial plane, which features additional centrifugal forces, is no 
more unlikely to feature outflows than polar regions if the star is 
assumed to remain electrically neutral.
\end{itemize}

    Thirdly, pseudo-periodic gamma ray bursts (GRBs) could be generated by 
a mechanism similar to the microquasar. Consider a rapidly rotating white 
dwarf with a companion star providing a steady supply of material to an 
orbiting accretion disk. Metastable oscillations could be established 
whereby the core of the white dwarf collapses to a neutron torus once 
sufficient matter and angular momentum has accumulated. This results in a 
brief period of jet activity in which enough mass and angular momentum is 
expelled to restore the star to a pure spheroidal white dwarf. Accretion 
of matter from the binary companion then repopulates the accretion disk, 
with the mass and angular momentum of the white dwarf slowly increasing 
until the cycle repeats. It may be that some of the shorter-duration gamma 
ray bursts can be attributed to situations like this.

\bigskip

\appendix{\centerline {\bf APPENDIX B} }

\renewcommand{\theequation}{B.\arabic{equation}}
\renewcommand{\thefigure}{B.\arabic{figure}}

\medskip
\noindent
    A year has elapsed since this paper was deposited on the archive. 
During that time,  explanations for the stability of toroidal black holes 
have not been ventured. An open discussion along these lines could have 
been included in the original submission but was not, partly to avoid 
polarising the views of a sympathetic audience and partly because of the 
bewildering array of possibilities. Though no claim is made of a 
satisfactory resolution, this final version devotes an appendix offering a 
`snapshot' of my thoughts on this issue which have benefited from twelve 
months of distilled cogitation. Parallels between the Kerr BH and 
Newtonian analogues comprising a self-gravitating annulus are explored. A 
simple quantized Newtonian model is used to investigate possible 
consequences in relativistic settings. The intention is not to denigrate 
the existing research on black hole spacetimes which has been the focus of 
much solemn effort by dedicated practitioners of general relativity, 
rather to communicate some pertinent concerns in a straightforward and 
hopefully thought-provoking manner.

    First it is shown that the ring singularity of the Kerr geometry 
travels with the velocity of light, regardless of the degree of rotation. 
Then it is demonstrated that a Newtonian equivalent of the Kerr geometry 
is a homogeneous self-gravitating ring of infinite density rotating at 
infinite velocity, implying an unphysically large angular momentum and 
kinetic energy. This is interpreted as an inevitable consequence of the 
simplistic model which entirely disregards the microscopic quantum nature 
of the ring. When a more realistic analysis is pursued in the Newtonian 
setting, it is observed that the velocity required for equilibrium is 
logarithmically relaxed - a macroscopically observable consequence of the 
quantum world. By implication it is then argued that a natural 
relativistic counterpart deviates from the Kerr solution, and that this 
deviation would be particularly evident at high angular momenta. 
Astrophysical environments of interest are then addressed to illustrate 
the possibility that event horizon topologies may not be restricted to 
2-spheres.

In pseudo-Cartesian coordinates $(\bar{t},x,y,z)$, the Kerr metric reads:
\begin{multline}
ds^2=d\bar{t}^2+dx^2+dy^2+dz^2+\frac{2mr^3}{r^4+a^2z^2}\\
\times\Bigl[d\bar{t}+\frac{(rx+ay)dx+(ry-ax)dy}{a^2+r^2}+\frac{z}{r}
dz\Bigr]^2
\end{multline}
and in Boyer-Lindquist coordinates $(t,r,\theta,\phi)$:
\begin{multline}
ds^2=-\frac{\Delta}{\rho^2}(a\sin^2\theta d\phi-dt)^2\\
+\frac{\sin^2\theta}{\rho^2}[(r^2+a^2)d\phi-adt]^2
+\frac{\rho^2}{\Delta}dr^2+\rho^2d\theta^2
\label{bl}
\end{multline}
\begin{align}
\Delta&=r^2-2mr+a^2\\
\rho^2&=r^2+a^2\cos^2\theta
\end{align}

The spatial coordinates of the two metrics obey the transformations:
\begin{align}
x&=r\sin\theta\cos\phi+a\sin\theta\sin\phi\\
y&=r\sin\theta\sin\phi-a\sin\theta\cos\phi\\
z&=r\cos\theta
\end{align}

    A true spherical polar coordinate $R$, coinciding asymptotically with 
$r$ at large radii, can be defined as $R^2=x^2+y^2+z^2$ which transforms 
to $R^2=r^2+a^2\sin^2\theta$. The ring singularity of the Kerr BH resides 
at $(r=0,\theta=\pi/2)$ or $(R=a,\theta=\pi/2)$. Because the Kerr solution 
is stationary, the circle on which the singularity lies is a geodesic. To 
investigate its velocity, it is assumed that $r$ and $\cos\theta$ are 
small, $r^2$ is negligible and $dr=d\theta=0$. The Boyer-Lindquist metric 
then becomes:
\begin{align}
ds^2=2mr\frac{(a\sin^2\theta d\phi-dt)^2}{a^2\cos^2\theta}
\end{align}

\begin{figure}
\centerline{\psfig{figure=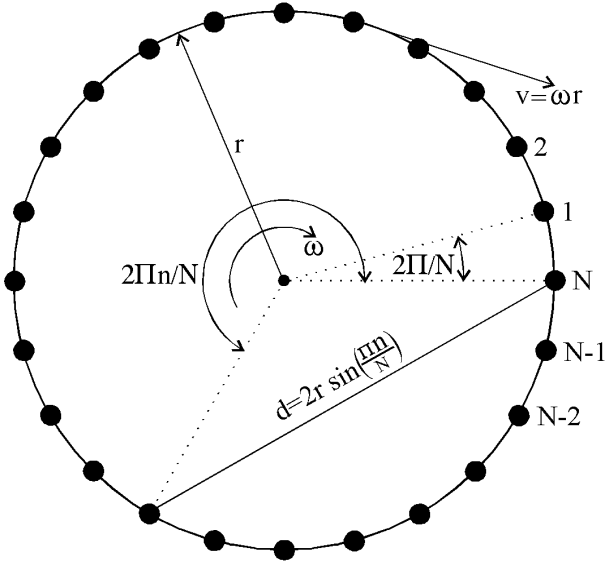,width=0.45\textwidth,angle=0}}
\caption{Self-gravitating ring of discrete particles.}
\end{figure}

    As $\cos\theta$ becomes infinitesimally small, to preserve $ds^2\leq 
1$, the numerator must converge as rapidly as $\cos^2\theta$ to zero, so 
the term within parentheses vanishes yielding
\begin{align}
v_{\rm ring}=R\frac{d\phi}{dt}=\frac{\sqrt{r^2+a^2\sin^2\theta}}{
a\sin^2\theta}=1
\end{align}
showing that the singularity travels with coordinate velocity $c$ when 
$a^2>0$. This is to be compared with the coordinate velocity of particles 
and photons remaining on the equator of the Kerr BH event horizon but 
being dragged around with the horizon. To investigate this, after setting 
$dr=d\theta=0$ and $\theta=\pi/2$ the metric reduces to:
\begin{align}
rds^2\!\!=\!
(r\!-2m)dt^2\!\!+\!4am\,dt\,d\phi\!-\!(r^3\!\!+\!ra^2\!\!+\!2ma^2)d\phi^2
\end{align}

To remain on the horizon, $ds^2$ is necessarily zero and the radius at the 
event horizon is given by $r_+=m+\sqrt{m^2-a^2}$ so that $R_+^{\,2}
=2m(m+\sqrt{m^2-a^2})=2mr_+$. Solving for $d\phi/dt$ allows the coordinate 
velocity of the equatorial event horizon to be determined:
\begin{align}
v_{\rm eeh}=R_+\frac{d\phi}{dt}\biggl\vert_{r_+}=\biggl[\frac{a^2}{
2m(m+\sqrt{m^2-a^2})}\biggr]^{1/2}
\end{align}
which varies from $v_{\rm eeh}=0.5a/m$ as $a\to 0$ to $v_{\rm eeh}
\sim0.7a/m$ as $a\to m$. The Kerr-Newman metric endowed with charge $Q$ 
features an outer event horizon at $r_+=M+\sqrt{m^2-Q^2-a^2}$ and, after 
restoring constants, remains sub-extremal if
\begin{align}
G^2m^2\geq GQ^2+c^2a^2
\end{align}
so charge and rotation act repulsively and in unison to drive the BH away 
from spherical symmetry. This conclusion seems inevitable without 
abandoning the equivalence principle (Petkov, 2001), which in any case 
invalidates the metric. Since extremality cannot be surpassed, general 
relativity restricts static particles to having $Q/M\leq\zeta_{\rm max}
=\sqrt{G}$. Gross violations are witnessed in particle physics, for 
protons $Q/M>10^{13}\zeta_{\rm max}$ and for electrons $Q/M>10^{16}\zeta_{
\rm max}$. Besides quantum scales, this means that a tiny electrical 
charge can profoundly affect large-scale spacetime geometries. For 
instance, if an imbalance of one electron/proton exists for every $\sim10^{
13}$ neutrons, the time dilation at the surface of a neutron star is 
almost entirely eliminated. Electromagnetism can thus play an important 
role in the stability of charged tori.

    Consider a planar constellation of $N$ identical satellites of total 
mass $M$ evenly and symmetrically distributed on a circle (Fig.~B.1). 
According to Newtonian mechanics, equilibrium is given by the balancing of 
gravitational and centripetal forces acting on each satellite such that 
the radius, $r$, of the circular orbits remains constant. All satellites 
rotate at constant angular velocity $\omega$ so that their individual 
velocities are $v=\omega r$. Equilibrium is attained when:
\begin{align}
\frac{GM}{4Nr^2}\sum_{n=1}^{N-1}\frac{1}{\sin(\frac{n\pi}{N})}=\frac{v^2}{
r}
\end{align}

In the limit $N\to\infty$, the series can be expressed in the form of a 
definite integral:
\begin{align}
\lim_{N\to\infty}\frac{1}{N}\sum_{n=1}^{N-1}\frac{1}{\sin(\frac{n\pi}{N})}
=\int_0^\pi\frac{dx}{\sin x}=\frac{4rv^2}{GM}
\end{align}

Although the series converges for $N<\infty$, the integral is divergent 
since
\begin{align}
\int_0^\pi\!\!\frac{dx}{\sin x}=\!\Biggl[\log\Bigl\vert\tan\frac{x}{2}
\Bigr\vert\Biggr]_0^\pi\!\!=2\times\lim_{x\to +0}\log\biggl(\frac{2}{x}
\biggr)
\end{align}

    Hence, a ring of finite total mass $M$ composed of an infinitude of 
individual satellites is obliged to rotate at infinite velocity in order 
to maintain radial equilibrium. This in turn implies an infinite angular 
momentum and kinetic energy for homogeneous rings. Planetary ring systems, 
such as Saturn's, are nevertheless observed. Despite their self-gravity, 
the rotational velocity remains modest -- evidence that their microscopic 
structure is particulate, not homogeneous. Where self-gravitating rings of 
matter are concerned, homogeneity is an unreasonable assumption, but one 
which is often adopted in general relativity as exemplified by the Kerr 
metric.

    Some of the following statements are necessarily speculative, but it 
should be kept in mind that the overly complacent alternative is to trust 
in mathematical models when conditions are far more extreme than those 
accessible to experimental tests. Acceleration of the Kerr singularity to 
the speed of light formally requires an infinite amount of energy. This 
energy must have been supplied by the gravitational potential energy of 
the matter which the black hole consumes. It is implausible that this 
energy source is truly unbounded, allowing the spacetime to attain 
infinite curvature. Rather, the collapse must halt at some limiting 
density, at which time the kinetic energy remains finite. Therefore, 
though the Newtonian analysis presented is very simplistic, aspects of 
this simplicity are also shared by its relativistic counterpart, the Kerr 
black hole.

    In an attempt to circumvent these limitations and prepare rough 
estimates, suppose that the ring structure is subdivided into a 
non-infinite number of satellites. For instance, the maximum number of 
neutrons contained in a mass of $10^9\msun$ is $\epsilon^{-1}=1.2\times10^{
66}$. In this case, the equilibrium satellite velocity remains finite with 
magnitude
\begin{align}
v\approx\sqrt{\frac{GM}{4r}\int_\epsilon^{\pi-\epsilon}\!\!\!\frac{dx}{
\sin x}}\sim 8.7\sqrt\frac{GM}{r}
\end{align}

    But in these situations, neutrons are plausibly replaced by 
Planck-scale particles. The characteristic Planck length is $l_{\rm p}
=\sqrt{\hbar G/c^3}\sim 1.6\times 10^{-35}$m and the Planck mass is $m_{
\rm p}=\sqrt{\hbar c/G}\sim 2.2\times 10^{-8}$kg. A Kerr BH of mass $M$ 
contains a  singularity of maximal radius when $a=M$ or $R=MG/c^2$ in 
natural units. Suppose that the singularity is a crystalline structure of 
Planck `particles' in a circular arrangement whose further collapse is 
resisted by quantum mechanical repulsion. One might object that this 
arrangement is too idealised or that it requires an infinite universal 
time to elapse - perhaps so, but it is here argued that some form of 
uncertainty principle or holographic correspondence prevents external 
observers from distinguishing between the present assumption and any other 
model. The mean separation between adjacent Planck particles is determined 
to be {\it independent of the total mass} and given by $d_{\rm sep}=2\pi 
Rm_{\rm p}/M=2\pi\sqrt{\hbar G/c^3}=2\pi l_{\rm p}$, a coincidence 
substantiating the original premise of a Planckian singularity. For a 
$10^9\msun$ BH, the total number of particles is $\epsilon^{-1}=9\times 
10^{46}$ which yields $v\sim 7.4\sqrt{GM/r}=7.4$c. The equivalent 
relativistic velocity is $\Gamma=8.4$ or $v_{\rm rel}\sim0.993$c (for a 
stellar mass BH, $v_{\rm rel}\sim0.992$c). It is known that the ratio 
$a/m$ for a Kerr BH can realistically approach $\sim0.998$ in 
astrophysical circumstances (Thorne, 1974), allowing scope for deviation 
from the Kerr geometry due to Planck scale phenomena.

    Though the margin for this to occur seems slender for isolated 
horizons, indicative of almost negligible deviation, the assumption that 
the singularity is quantum-mechanically sustained at the Planck scale 
introduces additional considerations which, in typical settings, are 
favourable for significant deformation of the Kerr geometry. The spacetime 
curvature in the vicinity of the singularity is tamed by the repulsion 
between Planck particles (e.g. Louko \& Matschull, (2001) where some 
success has been claimed regarding the quantization of geometry). This 
repulsion presumably grows as the ring's radius and $a/m$ decreases. 
Whereas the Kerr singularity is infinitely distant from the black hole's 
exterior due to the immensely strong curvature (Thorne, Price \& 
MacDonald, 1986), the Planckian singularity is susceptible to the 
influence of other matter through tidal gravity. Indeed, the surface 
gravity of a Kerr-Newman BH diminishes with increasing charge and 
rotation, disappearing altogether at extremality. So without infinite 
curvature, the singularity would otherwise be deformable.

    Most modern attempts at unification invoke extra spatial dimensions. 
At high energies, gravity is thought to `leak' away from our 3 dimensions. 
This has yet to be confirmed experimentally, due to practical 
difficulties. Measurements have verified the inverse square law down to 
submillimetre scales, equating to an energy scale $\sim10^{-2}$eV : well 
short of particle accelerator energies $\sim100$GeV, the sypersymmetry 
scale $\sim10^{16}$GeV and the Planck scale $m_{\rm p}c^2\sim10^{19}$GeV. 
Diminution of gravitational interactions at small scales would inevitably 
cause the orbital velocities of BH singularities to decrease further, 
quite feasibly by a substantial amount. General relativity will certainly 
break down at the highest energies, otherwise collapse cannot even be 
halted at the Planck scale. Since energy conditions are known to be 
violated by the Casimir effect and Hawking radiation, it is unlikely that 
they will be satisfied everywhere within a black hole. Conversely, the 
extreme pathology of closed timelike curves exhibited by the stationary BH 
metrics is tacitly embraced by investigators.

    Consider an accretion disk orbiting a rapidly rotating BH. The disk 
gravity induces tidal stresses on the Planckian singularity which tends to 
stretch it radially. In addition to the accretion disk, AGN reside within 
a molecular torus and host galaxy, each equatorially oriented with respect 
to the BH. In the context of rotary core collapse SNe, the outer shells 
collapse towards the plane of rotation where a substantial proportion of 
the debris forms a thick disk. Its mass and proximity to the core impose 
tidal forces whose influence on the central BH will be more pronounced 
than in AGN situations. The gravitational potential at radius $r$ in the 
plane of a circular hoop of radius $R\geq r\geq0$ whose linear mass 
density is $\lambda$ is given by the following function containing a 
complete elliptic integral of the first kind K(k):
\begin{align}
\Phi_{\rm hoop}(r)=\frac{R}{r}\Phi_{\rm hoop}\Bigl(\frac{R^2}{r}
\Bigr)=-\frac{4G\lambda R}{ R+r}K\biggl(\frac{2\sqrt{Rr}}{R+r}\biggr)
\end{align}

$\Phi_{\rm hoop}$ decreases monotonically from the centre ($r=0$) to 
negative infinity as $r\to R$. In elementary functions, it can be proven 
(by using series expansions for the elliptic integral and differentiating) 
that the internal ($g_{\rm\,int}$) and external ($g_{\rm\,ext}$) 
gravitational accelerations are bounded as follows:
\begin{align}
\pi\leq\frac{(R^{\,2}\!\!-r^2)g_{\rm\,int}}{rG\lambda}\leq4\leq \frac{
(r^2\!-R^{\,2})g_{\rm\,ext}}{RG\lambda}\leq2\pi
\end{align}

    The lower bound corresponds to $r\ll R$, the central limit to $r\to R$ 
and the upper bound is the asymptotic behaviour as $r\to \infty$. These 
functions are readily integrated in many situations and permit the simple 
construction of models in which one may be interested in calculating 
conservative estimates. For increased accuracy, interpolations are 
available whose maximum errors are $\sim0.2\%$ at $r/R\sim 0.57$ for the 
internal gravity and $\sim0.9\%$ at $r/R\sim 1.29$ for the external 
gravity, which is oppositely directed:
\begin{align}
g_{\rm\,int}(r)&\approx G\lambda r\biggl[\frac{4}{R^{\,2}-r^2}-\frac{4-\pi}
{R(R^{\,2}-r^2)^{1/2}}\biggr]\\
g_{\rm\,ext}(r)&\approx G\lambda R\biggl[\frac{4}{R^{\,2}-r^2}-\frac{
2\pi-4}{r^{3/2}(r^2-R^{\,2})^{1/4}}\biggr]
\end{align}

    Of immediate interest are truncated disks of constant areal mass 
densities $\sigma$. Integration leads to the following approximation for 
the gravity within a thin truncated disk ($0\leq r\leq R_\alpha<R_\beta$) 
where $R_\alpha$ and $R_\beta$ are the disk's radii at the inner and outer 
rims respectively:
\begin{multline}
g\approx 2G\sigma\log\biggl[\frac{(R_\alpha+r)(R_\beta-r)}{
(R_\alpha-r)(R_\beta+r)}\biggr]\\
-(4-\pi)G\sigma\biggl[\cos^{-1}\!\Bigl(\frac{r}{R_\beta}\Bigr)\!-\cos^{-1}
\!\Bigl(\frac{r}{ R_\alpha}\Bigr)\biggr]
\end{multline}

    It is often assumed that BH accretion disks are truncated some 
distance from the event horizon because stable circular orbits are 
forbidden within $r_{\rm ms}$, the innermost radius of marginal  
stability. For Schwarzschild BHs, $r_{\rm ms}=6M$, for retrograde extremal 
Kerr orbits, $r_{\rm ms}=9M$, and for prograde extremal Kerr orbits, $r_{
\rm ms}\to M$. Instability implies that material penetrating within these 
boundaries is accelerated towards the event horizon, so that these regions 
can scarcely be totally vacated. The acceleration experienced by infalling 
material, according to Newton's third law, is counterbalanced by a smaller 
but oppositely directed acceleration of the Planckian singularity towards 
the surrounding disk.  When the singularity is teased towards larger 
radii, the BH is likely to accumulate further angular momentum. Hence, 
continued accretion and tidal distortion can reinforce the distortion, 
making it conceivable that nonstationary accretion into the event horizon, 
coupled with continuous disk replenishment, could maintain a TBH against 
collapse for prolonged periods. According to this view, a TBH might evolve 
from a spheroidal BH and vice versa --- topological transitions of both 
kinds being possible. This enhances the probability of TBH formation, 
circumventing the need for collapse of a toroidal dust cloud or star as 
outlined in section~2. It also raises the question of whether the nuclei 
of some galaxies are revived into activity following a dormant stage e.g. 
by tidal interaction with nearby galaxies or more directly, following 
mergers. Elliptical galaxies, believed to be the outcomes of mergers, host 
an overabundance of AGN.

\begin{figure}
\centerline{\psfig{figure=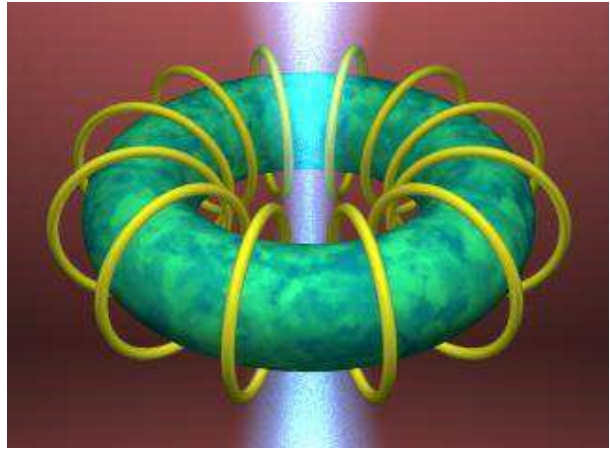,width=0.45\textwidth,angle=0}}
\caption{Depiction of a rotating electrically charged torus exhibiting a 
dipole magnetic field in the non-rotating frame. The field lines channel 
plasma from a surrounding accretion disk into the central aperture where 
jets are outwardly accelerated along the spin axis by ergoregional 
interactions.}
\end{figure}

    In conclusion, though Kerr's metric for rotating black holes is 
rightly celebrated as one of the landmark discoveries of 20th century 
science, the subsequent failure to satisfactorily unite the classical 
theory of general relativity with quantum mechanics remains a severe 
impediment to an understanding of nature. Until this fundamental obstacle 
can be overcome, one must be fully aware of the inherent drawbacks of 
myopic alternatives. Black hole topology is a subject unsuitable for 
purely classical computations, however sophisticated the mathematical 
veneer. Extreme caution should be employed if results are to be usefully 
construed. The rule of thumb calculations presented here are suggestive 
that an ultimate theory might accommodate the possibility of toroidal 
black holes, temporarily stabilised on astrophysically relevant timescales 
by the action of accretion, tidal deformation and Planck scale repulsion. 
The basis of the scenario advocated for supporting TBH stability is 
summarised below:

\begin{itemize}
\item
Black holes will often be formed with nearly maximal rotation e.g. from 
collapse of moderately rotating stellar progenitors. Accretion via a thin 
disk can also lead to a rapid build-up angular momentum. Astrophysically 
realistic ratios of $a/m$ can approach 0.998. Near-extremal black holes 
are therefore predicted to exist in rotary core-collapse supernovae, 
galactic nuclei and accreting black hole binary systems. In all 
situations, tidal stresses are exerted by matter exterior to the event 
horizon which orbits in the equatorial plane. Disk truncation decreases as 
extremality is approached, the radius of the innermost stable orbit and 
the singularity converge towards that of the event horizon. Mass steadily 
migrates into the event horizon by accretion, bridging the zone of disk 
truncation and bolstering the tidal forces acting on the singularity.
\item
The Kerr geometry is in some respects unphysical --- it contains a 
homogeneous ring singularity rotating at the speed of light, infinite 
spacetime curvature at the location of the singularity, closed timelike 
paths within its Cauchy horizon and makes no provision whatsoever for 
quantum mechanics. Furthermore, all attempts to match the exterior metric 
with realistic non-vacuum internal distributions of matter have thus far 
failed.
\item
If collapse halts at the Planck-scale, the velocity of rotation could 
decrease to $\sim0.993$c, suppressing the spacetime curvature in the 
vicinity of the singularity making it susceptible to tidal deformation by 
surrounding structures e.g. thin disks, thick disks and tori. High energy 
effects such as gravity `leaking' into higher dimensions may additionally 
modify the rotational velocity of the singularity. Radial elongation of 
the Planck singularity is conducive to the accumulation of angular 
momentum beyond that of undistorted configurations, acting as a barrier 
against collapse should accretion be interrupted. Sustained accretion and 
tidal distortion {\it might} ultimately result in toroidal event horizon 
topology.
\end{itemize}


\label{lastpage}

\end{document}